\RequirePackage[british]{babel}
\documentclass[reqno,a4paper,12pt,final]{amsart}
\usepackage[a4paper,hmargin=2cm,tmargin=3cm,bmargin=3cm]{geometry}
\usepackage[utf8x]{inputenc}
\usepackage{amsmath,amssymb,amstext,amsthm,amscd,mathrsfs,eucal}
\usepackage{graphicx,color}
\usepackage{array}
\usepackage[nosort]{cite}
\usepackage[vcentermath]{youngtab}
\Yboxdim{4pt}
\usepackage{hyperref}
\hypersetup{%
  pdftitle   = {On the Lie-algebraic origin of metric 3-algebras},
  pdfkeywords = {M2, M Theory, Bagger-Lambert, Lie algebra, Filippov, 3-algebra, Lie triple system},
  pdfauthor  = {Paul de Medeiros, José Figueroa-O'Farrill, Elena Méndez-Escobar, Patricia Ritter},
  pdfcreator = {\LaTeX\ with package \flqq hyperref\frqq}
}
\PrerenderUnicode{éÉ}
\newcommand{\fg}{\mathfrak{g}}

\newcommand{\fgl}{\mathfrak{gl}}
\newcommand{\fh}{\mathfrak{h}}

\newcommand{\fk}{\mathfrak{k}}

\newcommand{\fs}{\mathfrak{s}}

\newcommand{\fso}{\mathfrak{so}}

\newcommand{\fsp}{\mathfrak{sp}}
\newcommand{\fusp}{\mathfrak{usp}}
\newcommand{\fsu}{\mathfrak{su}}
\newcommand{\fu}{\mathfrak{u}}

\newcommand{\SU}{\mathrm{SU}}
\newcommand{\Sp}{\mathrm{Sp}}
\newcommand{\U}{\mathrm{U}}
\newcommand{\SO}{\mathrm{SO}}

\newcommand{\RR}{\mathbb{R}}
\newcommand{\CC}{\mathbb{C}}
\newcommand{\DD}{\mathbb{D}}
\newcommand{\OO}{\mathbb{O}}
\newcommand{\ZZ}{\mathbb{Z}}
\newcommand{\XX}{\mathbb{X}}
\newcommand{\YY}{\mathbb{Y}}
\newcommand{\XXbar}{\overline{\mathbb{X}}}

\newcommand{\eC}{\mathscr{C}}
\newcommand{\eD}{\mathscr{D}}

\newcommand{\be}{\boldsymbol{e}}

\newcommand{\Vbar}{\overline{V}}
\renewcommand{\Im}{\mathrm{Im}}
\renewcommand{\Re}{\mathrm{Re}}

\DeclareMathOperator{\End}{End}

\DeclareMathOperator{\Mat}{Mat}

\DeclareMathOperator{\ad}{ad}

\DeclareMathOperator{\id}{id}

\DeclareMathOperator{\im}{im}
\DeclareMathOperator{\pr}{pr}
\DeclareMathOperator{\tr}{tr}
\DeclareMathOperator{\Tr}{Tr}

\newcommand{\hfl}[1]{\overline{#1}^\flat}
\newcommand{\hsh}[1]{\overline{#1}^\sharp}
\newcommand{\rf}[1]{[\![#1]\!]}
\theoremstyle{plain}
\newtheorem{lemma}{Lemma}
\newtheorem{proposition}[lemma]{Proposition}
\newtheorem{theorem}[lemma]{Theorem}

\theoremstyle{definition}
\newtheorem{definition}[lemma]{Definition}
\newtheorem{remark}[lemma]{Remark}
\newtheorem{example}[lemma]{Example}
\newcommand{\MUNCH}[1]{\relax}

\allowdisplaybreaks
\begin{document}
\title{On the Lie-algebraic origin of metric 3-algebras}
\author[de Medeiros, Figueroa-O'Farrill, Méndez-Escobar, Ritter]{Paul de Medeiros, José Figueroa-O'Farrill, Elena Méndez-Escobar and Patricia Ritter}
\address{School of Mathematics and Maxwell Institute for Mathematical Sciences, University of Edinburgh, James Clerk Maxwell Building, King's Buildings,
  Edinburgh EH9 3JZ, UK}
\address[JMF also]{Departament de Física Teòrica \& IFIC (CSIC-UVEG), Universitat de València, 46100 Burjassot, Spain}
\email{\{P.deMedeiros,J.M.Figueroa,E.Mendez\}@ed.ac.uk, P.D.Ritter@sms.ed.ac.uk}
%\date{\today}
\begin{abstract}
  Since the pioneering work of Bagger--Lambert and Gustavsson, there has been a proliferation of three-dimensional superconformal Chern--Simons theories
  whose main ingredient is a metric 3-algebra.  On the other hand, many of these theories have been shown to allow for a reformulation in terms of
  standard gauge theory coupled to matter, where the 3-algebra does not appear explicitly.  In this paper we reconcile these two sets of results by
  pointing out the Lie-algebraic origin of some metric 3-algebras, including those which have already appeared in three-dimensional superconformal
  Chern--Simons theories.  More precisely, we show that the real 3-algebras of Cherkis--Sämann, which include the metric Lie 3-algebras as a special
  case, and the hermitian 3-algebras of Bagger--Lambert can be constructed from pairs consisting of a metric real Lie algebra and a faithful (real or
  complex, respectively) unitary representation.  This construction generalises and we will see how to construct many kinds of metric 3-algebras from
  pairs consisting of a real metric Lie algebra and a faithful (real, complex or quaternionic) unitary representation.  In the real case, these
  3-algebras are precisely the Cherkis--Sämann algebras, which are then completely characterised in terms of this data.  In the complex and quaternionic
  cases, they constitute generalisations of the Bagger--Lambert hermitian 3-algebras and anti-Lie triple systems, respectively, which underlie $N{=}6$
  and $N{=}5$ superconformal Chern--Simons theories, respectively.  In the process we rederive the relation between certain types of complex 3-algebras
  and metric Lie superalgebras.
\end{abstract}
\maketitle
\tableofcontents

\section{Contextualisation and introduction}
\label{sec:introduction}

The foundational work of Bagger and Lambert \cite{BL1} and Gustavsson \cite{GustavssonAlgM2}, culminating in the theory proposed in \cite{BL2}, has
prompted a great deal of progress recently \cite{GaiottoWitten, pre3Lee, MaldacenaBL, KlebanovBL, MauriPetkou, 3Lee, BL4, CherSaem, SchnablTachikawa,
  ABJ, OoguriPark, JT, BergshoeffMM2, BdRHR-BL, BHRSS} in our ability to construct new superconformal field theories in three dimensions that are
thought to describe the low-energy dynamics of configurations of multiple coincident M2-branes in M-theory.  These new superconformal field theories all
involve a non-dynamical gauge field, described by a Chern--Simons-like term in the lagrangian, which is coupled to matter fields parametrising the
degrees of freedom transverse to the worldvolume of the M2-branes.  The order of the matter couplings in these theories is dictated by superconformal
invariance and indeed several other features are as expected (see \cite{SchwarzCS,GYCS} and references therein) following analyses based on existing
$N{=}1$ and $N{=}2$ superfield techniques in three dimensions.

The novelty in the proposal of \cite{BL1,GustavssonAlgM2,BL2} for a theory of this kind with maximal $N{=}8$ supersymmetry is that the interactions
involve an algebraic object known as a Lie 3-algebra, instead of the Lie algebras commonly used to describe conventional gauge couplings.  A lagrangian
description of this theory requires the Lie 3-algebra to admit an ``invariant'' inner product.  (See, for example, \cite[Remark~8]{Lor3Lie}.)  Demanding
the signature of this inner product to be positive-definite ensures the unitarity of the resulting quantum theory, but restricts
\cite{NagykLie,GP3Lie,GG3Lie} one for all practical purposes to the unique indecomposable example already considered in \cite{BL2}.

To some extent, the Bagger-Lambert theory in \cite{BL2} has now been subsumed into an $N{=}6$ superconformal Chern--Simons-matter theory proposed by
Aharony, Bergman, Jafferis and Maldacena \cite{MaldacenaBL} (see also \cite{KlebanovBL}).  This $N{=}6$ theory involves ordinary gauge couplings for
matter fields valued in the bifundamental representation of the Lie algebra $\fu(n) \oplus \fu(n)$ (or $\fsu(n) \oplus \fsu(n)$ via a certain truncation) and coincides with \cite{BL2} for the special case
of $n{=}2$, where supersymmetry is enhanced to $N{=}8$.  Generically it is thought to describe $n$ coincident M2-branes on the orbifold $\RR^8 / \ZZ_k$,
where $k$ denotes the integer level of the Chern--Simons term.  Despite the conventional appearance of the gauge symmetry in this theory, Bagger and
Lambert \cite{BL4} subsequently showed that it too can be recast in terms of a metric 3-algebra, albeit of a different kind from the metric Lie
3-algebras of \cite{BL2}.

Another type of superconformal field theory which, like \cite{MaldacenaBL,BL4}, has an $\SU(4) \times \U(1)$ global symmetry but is generically only
$N{=}2$ supersymmetric has been proposed by Cherkis and Sämann \cite{CherSaem}.  This theory is also based on a metric 3-algebra, albeit different from
the metric Lie 3-algebras of \cite{BL2}, which they generalise, and from the metric 3-algebras of \cite{BL4}.

Generalisations of the model in \cite{MaldacenaBL} have been considered in \cite{3Lee,SchnablTachikawa,BHRSS,ABJ}, using a plethora of different
techniques, though all result in conventional Chern--Simons theories coupled to matter fields valued in the bifundamental or similar tensor-product
representation of a variety of possible Lie algebras.  With generic $N{=}6$ supersymmetry, it has been found \cite{3Lee,SchnablTachikawa,BHRSS} that
this Lie algebra can be either $\fsu(m) \oplus \fsu(n) \oplus \fu(1)$ or $\fsp(m) \oplus \fu(1)$ (up to additional factors of $\fu(1)$).  (In a revision
to \cite{BL4}, Bagger and Lambert have also written the 3-algebra that corresponds to the $\fu(m) \oplus \fu(n)$ case.)  With generic $N{=}5$
supersymmetry, the Lie algebra can be either $\fso(m) \oplus \fsp(n)$ \cite{3Lee}, $\fso(7) \oplus \fsp(1)$, $\fg_2 \oplus \fsp(1)$ or $\fso(4) \oplus
\fsp(1)$ \cite{BHRSS}.  For the regular cases where $m \neq n$, these theories have been interpreted \cite{ABJ} in terms of $|m - n|$ fractional
M2-branes probing the singularity of the M-theory background $\RR^8 / \ZZ_k$ in the unitary case and $\RR^8 / {\hat D}_k$ in the orthogonal and
symplectic cases (${\hat D}_k$ being the binary dihedral group of order $4k$).  Consistent $N{=}4$ truncations of all these new theories exist
\cite{BHRSS} from which one can recover some previously known classes of $N{=}4$ superconformal Chern--Simons-matter theories
\cite{GaiottoWitten,pre3Lee}.  There have also been new proposals for the holographic duals of some smooth M-theory backgrounds of the form $AdS_4
\times X_7$ which are less than half-BPS, and again they involve conventional superconformal Chern--Simons-matter theories with quiver gauge groups.
The $N{=}3$ superconformal theories dual to backgrounds for which $X_7$ is a 3-Sasaki manifold arising from a particular quotient construction have been
described in \cite{JT}.  An $N{=}1$ truncation of the model in \cite{MaldacenaBL}, obtained by breaking the $\SU(4)$ R-symmetry to $\Sp(2)$, has been
given in \cite{OoguriPark} which corresponds to $X_7$ being a squashed $S^7$.

All this prompts one to question whether the 3-algebras appearing in the constructions \cite{BL1,GustavssonAlgM2,BL2,BL4,CherSaem} play a fundamental
rôle in M-theory or, at least insofar as the effective field theory is concerned, are largely superfluous.  The equivalence of \cite{BL4} and
\cite{MaldacenaBL} and the abundance of new theories (dual to known M-theory backgrounds) which seem not to involve a 3-algebra might suggest the
latter.  Nonetheless, given our lack of understanding of how to incorporate in Lie-algebraic terms the expected properties of M-theoretic degrees of
freedom, like the entropy scaling laws for M2- and M5-brane condensates, it may be useful to understand the precise relation between the 3-algebras
appearing in the recent literature on superconformal Chern--Simons theory and Lie algebras.

In this paper we will do precisely this.  We will show how certain types of metric 3-algebras, which include the ones which have appeared in the recent
literature, can be deconstructed into and reconstructed from pairs consisting of a metric real Lie algebra and a faithful unitary representation.  In a
way this reduces the problem of constructing and even characterising such metric 3-algebras to the problem of classifying metric Lie subalgebras of the
orthogonal, unitary and unitary symplectic Lie algebras.  Our results take their inspiration from a general algebraic construction of pairs due to
Faulkner \cite{FaulknerIdeals}, which we will review below; although for pedagogical reasons we have decided not to present the results in this paper as
a series of specialisations of Faulkner's construction, but rather to present this construction as an \emph{a posteriori} unifying theme.  In a
companion paper we will present the construction of superconformal Chern--Simons theories from the Lie-algebraic datum --- one principal aim being to
translate physical properties of the field theory into Lie representation theory.

This paper is organised as follows.  In Section \ref{sec:CS} we review the definition of the generalised metric Lie 3-algebras of \cite{CherSaem} and
extract from every such 3-algebra a pair $(\fg,V)$ consisting of a metric real Lie algebra $\fg$ and a faithful real orthogonal representation $V$.  We
illustrate this with several examples: the unique nonabelian indecomposable euclidean Lie 3-algebra, the nonsimple nonabelian indecomposable lorentzian
Lie 3-algebras, the $\eC_{2d}$ 3-algebras of \cite{CherSaem} and a mild generalisation we denote $\eC_{m+n}$.  We then show how starting from such a
pair $(\fg,V)$ one can reconstruct a metric Lie 3-algebra in the class discussed in \cite{CherSaem} and in this way establish a one-to-one
correspondence between isomorphism classes of generalised metric Lie 3-algebras and isomorphism classes of pairs $(\fg,V)$, for natural notions of
isomorphism which will be defined.  This reduces the classification of generalised metric Lie 3-algebras to that of metric Lie subalgebras of $\fso(V)$.

In Section \ref{sec:BL4} we review the definition of the hermitian 3-algebras of \cite{BL4} and extract from every such 3-algebra a pair $(\fg,V)$
consisting again of a metric real Lie algebra $\fg$ and now a faithful complex unitary representation $V$.  We illustrate this with the examples of the
hermitian 3-algebras in (the revised version of) \cite{BL4}.  Conversely, starting from such a pair $(\fg,V)$ we show how to reconstruct a hermitian
3-algebra which in general will belong to a class which includes the 3-algebras of \cite{BL4} as special cases.  This then poses the problem of
recognising the 3-algebras of \cite{BL4} within this more general class.  We solve this problem by showing that these are precisely those hermitian
3-algebras which can be embedded in a complex metric Lie superalgebra with even subalgebra the complexification $\fg_\CC$ of $\fg$ and even subspace $V
\oplus V^*$.  This complex Lie superalgebra is itself the complexification of a real Lie superalgebra with underlying vector superspace $\fg \oplus
\rf{V}$, where $\rf{V}$ is the real representation of $\fg$ whose complexification is $V \oplus V^*$.  This result may be thought of as a 3-algebraic
rederivation of the relation between $N{=}6$ Chern--Simons theories and Lie superalgebras \cite{3Lee,SchnablTachikawa} based on earlier work on $N{=}4$
theories \cite{GaiottoWitten,pre3Lee} and sheds some light on the relationship between that work and \cite{BL4}.

In Section \ref{sec:generalisation} we introduce Faulkner's construction of pairs from a metric Lie algebra and a faithful representation and we show
how the constructions in previous sections correspond to the special cases where the representation is real orthogonal and complex unitary,
respectively.  The former construction appears already in \cite{FaulknerIdeals}, whereas the latter appears to be new.  For completeness, we also
present the construction of 3-algebras from quaternionic unitary representations, that being the only other type of inner product which admits a notion
of positive-definiteness.  The construction can be repeated for other types of inner products and some of them have been considered already in
\cite{FaulknerIdeals}, but we restrict ourselves in this paper to those which are likely to play a rôle in the construction of superconformal
Chern--Simons theories.  This case contains the metric 3-algebras which can be used to reformulate $N{=}5$ superconformal Chern--Simons theories.

Finally in Section \ref{sec:conclusions} we summarise the main results of this paper and comment on future directions of research that these results
suggest.

\section{The generalised metric Lie 3-algebras of Cherkis--Sämann}
\label{sec:CS}

In \cite{CherSaem} Cherkis and Sämann constructed three-dimensional $N{=}2$ superconformal Chern--Simons theories from a 3-algebra defined as follows
\cite[§4.1]{CherSaem}.

\begin{definition}\label{def:CS}
  A \textbf{generalised metric Lie 3-algebra} consists of a real inner product space $(V,\left<-,-\right>)$ and a trilinear bracket $\Phi: V \times V
  \times V \to V$, denoted $(x,y,z) \mapsto [x,y,z]$, satisfying the following axioms for all $x,y,z,v,w\in V$:
  \begin{enumerate}
  \item[(CS1)] the \emph{unitarity} condition
    \begin{equation}
      \label{eq:unitarity-CS}
      \left<[x,y,z], w\right> =  - \left<z, [x,y,w]\right>~;
    \end{equation}
  \item[(CS2)] the \emph{symmetry} condition
    \begin{equation}
      \label{eq:symmetry-CS}
      \left<[x,y,z], w\right> = \left<[z,w,x], y\right>~;
    \end{equation}
  \item[(CS3)] and the \emph{fundamental identity}
    \begin{equation}
      \label{eq:FI-CS}
      [x,y,[v,w,z]] - [v,w,[x,y,z]] = [[x,y,v],w,z] + [v,[x,y,w],z]~.
    \end{equation}
  \end{enumerate}
\end{definition}

We remark that axioms \eqref{eq:unitarity-CS} and \eqref{eq:symmetry-CS} together imply that $[x,y,z] = -[y,x,z]$, whence the bracket defines a
linear map $\Lambda^2 V \otimes V \to V$.  Decomposing $\Lambda^2 V \otimes V$ into
\begin{equation*}
  \Lambda^2 V \otimes V = \Lambda^3 V \oplus V^{\yng(2,1)}~,
\end{equation*}
where $V^{\yng(2,1)}$ contains the contraction with the inner product and is therefore not irreducible, breaks $\Phi$ up in turn into two components:
\begin{enumerate}
\item $\Phi_F : \Lambda^3V \to V$, which is totally skewsymmetric; and
\item $\Phi_L : V^{\yng(2,1)} \to V$, which is such that
  \begin{equation*}
    \Phi_L(x,y,z) + \Phi_L(z,x,y) + \Phi_L(y,z,x) = 0~.
  \end{equation*}
\end{enumerate}
There are thus two extremal cases of these 3-algebras:
\begin{itemize}
\item \textbf{(metric) Filippov} or \textbf{Lie 3-algebras}, where $\Phi=\Phi_F$, and
\item \textbf{(metric) Lie triple systems}, where $\Phi=\Phi_L$.
\end{itemize}
The former case coincides with the metric Lie 3-algebras of the $N{=}8$ theories \cite{BL1,GustavssonAlgM2,BL2}.  The general case, however, is a mixture
of these two.

We will now show that generalised metric Lie 3-algebras are in one-to-one correspondence with pairs $(\fg,V)$ consisting of a (real) metric Lie algebra
$\fg$ and a faithful orthogonal representation $V$, or equivalently, with metric subalgebras $\fg < \fso(V)$.  This is proved in two steps.  Given such
a 3-algebra $V$ we will associate a metric Lie algebra $\fg$, and conversely we will reconstruct such a 3-algebra from a pair $(\fg,V)$.

\subsection{Deconstruction}
\label{sec:deconstruction-CS}

Let $V$ be a generalised metric Lie 3-algebra and let us define a bilinear map $D: V \times V \to \End V$ by $D(x,y) z = [x,y,z]$.  We have seen before
that $D(x,y) = - D(y,x)$.  Moreover the unitarity axiom \eqref{eq:unitarity-CS}, when written in terms of $D$, becomes
\begin{equation*}
  \left<D(x,y) z, w\right> = - \left<z, D(x,y) w\right>~,
\end{equation*}
whence $D(x,y) \in \fso(V)$ is skewsymmetric.  Let $\fg$ denote the span of the $D(x,y)$ in $\End V$.  As we now show, it is closed under commutators.

\begin{proposition}
  Let $\fg = \im D$.  Then $\fg < \fso(V)$ is a Lie subalgebra.
\end{proposition}

\begin{proof}
  This is a consequence of the fundamental identity~\eqref{eq:FI-CS}.  Indeed, written in terms of $D$, the fundamental identity reads
  \begin{equation*}
    D(x,y) D(v,w) z - D(v,w) D(x,y) z = D(D(x,y) v, w) z + D(v, D(x,y) w) z~,
  \end{equation*}
  which upon abstracting $z$ can be rewritten as
  \begin{equation}
    \label{eq:DD-CS}
    [D(x,y), D(v,w)] = D(D(x,y) v, w) + D(v, D(x,y) w)~,
  \end{equation}
  which shows that $[\fg,\fg] \subset \fg$.
\end{proof}

Moreover we claim that $\fg$ is a metric Lie algebra, so that it possesses an ad-invariant inner product denoted by $\left(-,-\right)$.  This inner
product is defined by extending
\begin{equation}
  \label{eq:ip-CS}
  \left(D(x,y), D(z,w)\right) = \left<D(x,y)z, w\right>
\end{equation}
bilinearly to all of $\fg$.

\begin{proposition}
  The bilinear form on $\fg$ defined by \eqref{eq:ip-CS} is symmetric, nondegenerate and ad-invariant.
\end{proposition}

\begin{proof}
  The symmetry of the bilinear form \eqref{eq:ip-CS} is precisely the symmetry axiom \eqref{eq:symmetry-CS}.  Indeed, the symmetry axiom says that
  \begin{equation*}
    \left<[x,y,z], w\right> = \left<D(x,y)z,w\right> = \left(D(x,y),D(z,w)\right)
  \end{equation*}
  is equal to
  \begin{equation*}
    \left<[z,w,x], y\right> = \left<D(z,w)x,y\right> = \left(D(z,w),D(x,y)\right)~.
  \end{equation*}

  To prove nondegeneracy, let $\delta = \sum_i D(x_i,y_i)$ be such that $\left(\delta, D(u,v)\right) =0$ for all $u,v\in V$.  This means that
  \begin{equation*}
    \left<\delta u, v\right> = 0\quad\forall u,v\in V~.
  \end{equation*}
  Since the inner product on $V$ is nondegenerate, this means that $\delta u = 0$ for all $u\in V$, whence the endomorphism $\delta = 0$.  This means
  that \eqref{eq:ip-CS} does define an inner product.

  Finally, we prove the ad-invariance of the inner product:
  \begin{align*}
    \left(D(u,v),[D(w,x),D(y,z)]\right) &= \left(D(u,v),D(D(w,x) y,z) + D(y,D(w,x)z)\right) && \text{by \eqref{eq:DD-CS}}\\
    &= \left<D(u,v) D(w,x) y, z\right> + \left<D(u,v) y, D(w,x)z\right> && \text{by \eqref{eq:ip-CS}}\\
    &= \left<D(u,v) D(w,x) y, z\right> - \left<D(w,x) D(u,v) y, z\right> && \text{by \eqref{eq:unitarity-CS}}\\
    &= \left<[D(u,v), D(w,x)] y,z\right>\\
    &= \left([D(u,v), D(w,x)], D(y,z)\right) && \text{again by \eqref{eq:ip-CS}}~.
  \end{align*}
\end{proof}

It is worth stressing that the bilinear form $(-,-)$ in \eqref{eq:ip-CS} depends on the inner product $\left< -,- \right>$ on $V$ and is distinct from
the Killing form on $\fg$.  For the case of metric Lie 3-algebras, the distinction between these two objects has been noted already by Gustavsson in
\cite{Gustavsson1loop}.  It is the bilinear form in \eqref{eq:ip-CS} which appears in the Chern-Simons term in the superconformal field theories
associated with these 3-algebras, rather than the Killing on $\fg$ that is sometimes assumed.

Let us illustrate the aforementioned distinction with some examples.

\begin{example}\label{eg:A4}
  First of all we consider the original euclidean Lie 3-algebra in \cite{BL2}, denoted $A_4$.  The underlying vector space is $\RR^4$ with the standard
  euclidean structure.  Relative to an orthonormal basis $(e_1,\dots,e_4)$, the 3-bracket is given by
  \begin{equation}
    [e_i,e_j,e_k] =  \sum^{4}_{\ell=1} \epsilon_{ijk\ell} e_\ell~.
  \end{equation}
  The Lie algebra $\fg = \im D$ is spanned by generators $D_{ij} := D(e_i, e_j)$, for $i<j$.  It is not hard to see that these six generators are
  linearly independent, whence they must span all of $\fso(4)$, which is the gauge algebra of the original Bagger--Lambert model.
  Unlike the Killing form on $\fso(4)$, the inner product $(-,-)$ induced by the 3-algebra structure
  \begin{equation}
    \left( D_{ij}, D_{k\ell} \right) =  \left< [e_i, e_j, e_k], e_\ell \right>
  \end{equation}
  has indefinite signature. Indeed, for the six generators $D_{ij}$, the only nonzero inner products are
  \begin{equation}
    \left( D_{12}, D_{34} \right) = \left( D_{14}, D_{23} \right) = \left( D_{13}, D_{42} \right) = 1~,
  \end{equation}
  whence the signature is split.  This manifests itself in the fact that the levels of the Chern--Simons terms coming from the different $\fsu(2)$
  factors in $\fso(4)$ have opposite signs, as seen, for example, in \cite{VanRaamsdonkBL}.
\end{example}

The fact that the induced inner product on $\fg$ is not positive-definite is actually always the case for metric Lie 3-algebras.  Indeed, we have the
following

\begin{proposition}\label{pr:D-null}
  The $D(x,y)\in\fg$ have zero norm for all $x,y\in V$ if and only if the 3-bracket is totally skewsymmetric.
\end{proposition}

\begin{proof}
  Skewsymmetry of the bracket says that for all $x,y\in V$,
  \begin{equation*}
    \left(D(x,y),D(x,y)\right) = \left<[x,y,x],y\right> = 0~,
  \end{equation*}
  whence $D(x,y)$ has zero norm. (This property has been noted in equation (28) of \cite{Gustavsson1loop}.) Conversely, suppose that for all $x,y\in V$,
  $D(x,y)$ is null.  Polarising
  \begin{align*}
    0 &= \left(D(x,y+z), D(x,y+z)\right)\\
    &= \left(D(x,y) + D(x,z), D(x,y) + D(x,z)\right)\\
    &= 2 \left(D(x,y), D(x,z)\right)~.
  \end{align*}
  This says that $\left(D(x,y), D(z,w)\right) = \left<[x,y,z],w\right>$ is skewsymmetric under $x \leftrightarrow z$.  Since it is already skewsymmetric
  in $x \leftrightarrow y$ and $z \leftrightarrow w$, it follows that it is totally skewsymmetric.
\end{proof}

This means that in the case of a metric Lie 3-algebra, $\fg$ has a basis consisting of vectors of zero norm.  This means that the inner product on $\fg$
has split signature and in particular that $\fg$ is even-dimensional.

We continue with the lorentzian metric Lie 3-algebras of \cite{GMRBL,BRGTV,HIM-M2toD2rev,Lor3Lie}.

\begin{example}\label{eg:lor3lie}
  Let $\fs$ be a semisimple Lie algebra with a choice of ad-invariant inner product.  We let $W(\fs)$ denote the lorentzian Lie 3-algebra with underlying
  vector space $V = \fs \oplus \RR u \oplus \RR v$ with inner product extending the one on $\fs$ by declaring $u,v \perp \fs$ and $\left<u,v\right> = 1$
  and $\left<v,v\right>=0=\left<u,u\right>$, the last condition being a choice.  The nonzero 3-brackets are given by
  \begin{equation*}
    [u,x,y] = [x,y] \qquad\text{and}\qquad [x,y,z] = - \left<[x,y],z\right> v~,
  \end{equation*}
  for all $x,y,z\in\fs$.  The Lie algebra $\fg$ is the Lie algebra of inner derivations of the Lie 3-algebra, which was termed $\ad V$ in
  \cite{Lor3Lie}.  As shown for example in that paper, this algebra is isomorphic to $\fs \ltimes \fs_{\text{ab}}$ with generators $A_x=D(x_1,x_2)$ and
  $B_x=D(u,x)$ for $x\in\fs$ and where in the definition of $A_x$, we have $x=[x_1,x_2]$.  Since $\fs$ is semisimple, $[\fs,\fs]=\fs$ and hence such
  $x_1,x_2$ can always be found and moreover $A_x$ is independent of the precise choice of $x_1,x_2$ solving $x=[x_1,x_2]$.  The nonzero Lie brackets of
  $\fg$ are given by
  \begin{equation*}
    [A_x,A_y] = A_{[x,y]} \qquad\text{and}\qquad [A_x, B_y] = B_{[x,y]}~.
  \end{equation*}
  The inner product is then given by
  \begin{align*}
    \left(A_x,A_y\right) &= \left<[x_1,x_2,y_1],y_2\right> = - \left<x,y_1\right> \left<v,y_2\right> = 0\\
    \left(A_x,B_y\right) &= \left<[x_1,x_2,u],y\right> = \left<x,y\right>\\
    \left(B_x,B_y\right) &= \left<[u,x,u],y\right> = 0~,
  \end{align*}
  whence we see that the inner product is again split, as expected.
\end{example}

Next, we consider the $\eC_{2d}$ 3-algebras in \cite{CherSaem}.  These are not metric Lie 3-algebras.

\begin{example}\label{eg:CS}
  The underlying vector space $V$ of $\eC_{2d}$ is the real vector space of off-diagonal hermitian $2d \times 2d$ matrices:
  \begin{equation*}
    V = \left\{\begin{pmatrix} 0 & A \\ A^* & 0 \end{pmatrix} \middle | A \in \Mat_d(\CC)  \right\}~,
  \end{equation*}
  with $A^*$ the hermitian adjoint and with scalar product given by the trace of the product, which agrees with (twice) the real part of the natural
  hermitian inner product on the space of complex $d\times d$ matrices:
  \begin{equation*}
    \left<\begin{pmatrix} 0 & A \\ A^* & 0 \end{pmatrix}, \begin{pmatrix} 0 & B \\ B^* & 0 \end{pmatrix}\right> = \tr (AB^* + A^*B) = 2 \Re \tr A^*B~.
  \end{equation*}
  The 3-bracket is defined by $[x,y,z] := [[x,y]\tau,z]$, where $\tau = \begin{pmatrix} 1 & 0 \\ 0 & -1\end{pmatrix}$, for all $x,y,z\in V$ and where
  $[-,-]$ is the matrix commutator.  Letting $x = \begin{pmatrix} 0 & A \\ A^* & 0\end{pmatrix}$, $y = \begin{pmatrix} 0 & B \\ B^* & 0\end{pmatrix}$
  and $z = \begin{pmatrix} 0 & C \\ C^* & 0\end{pmatrix}$, a calculation reveals that
  \begin{equation*}
    [x,y,z] = 
    \begin{pmatrix}
      0 & (A B^* - B A^*) C + C (A^* B - B^* A) \\
      (B^* A - A^* B) C^* + C^* (B A^* - A B^*) & 0
    \end{pmatrix}~,
  \end{equation*}
  whence the action of $D(x,y)$ on $z$ is induced from the action of $D(x,y)$ on $C$, which is a linear combination of
  left multiplication by the skewhermitian $d\times d$ matrix $A B^* - B A^*$ and right multiplication by
  the skewhermitian $d\times d$ matrix $A^* B - B^* A$.  Let us decompose them into traceless + scalar matrices as follows
  \begin{equation*}
    A B^* - B A^* = S(A,B) + \tfrac{i}{d}\alpha(A,B) 1_d \qquad\text{and}\qquad
    A^* B - B^* A = S(A^*,B^*) - \tfrac{i}{d}\alpha(A,B) 1_d~, 
  \end{equation*}
  where $S(A,B)$ and $S(A^*,B^*)$ are traceless, whence in $\fsu(d)$, and $\alpha(A,B) = 2 \Im\tr A B^* \in \RR$.  Notice as well that
  \begin{equation*}
    (A B^* - B A^*) C + C (A^* B - B^* A)  = S(A,B) C + C S(A^*,B^*)~,
  \end{equation*}
  whence the Lie algebra $\fg$ is isomorphic to $\fsu(d) \oplus \fsu(d)$, acting on $V$ as the underlying real representation of
  $(\boldsymbol{d},\overline{\boldsymbol{d}}) \oplus (\overline{\boldsymbol{d}},\boldsymbol{d})$, which we denote
  $\rf{(\boldsymbol{d},\overline{\boldsymbol{d}})}$.  In other words, to the generalised metric Lie 3-algebra $\eC_{2d}$ one can associate the pair
  $(\fsu(d) \oplus \fsu(d), \rf{(\boldsymbol{d},\overline{\boldsymbol{d}})})$.  Notice finally that the invariant inner product on $\fsu(d) \oplus
  \fsu(d)$ has split signature, being given by $\left(X_L\oplus X_R, Y_L\oplus Y_R\right) = \tr (X_L Y_L - X_RY_R)$, for $X_L,Y_L,X_R,Y_R \in \fsu(d)$.

  It is possible to modify this example in order to construct 3-algebras denoted $\eC_{m+n}$, where the underlying vector space is the space of
  hermitian $(m+n) \times (m+n)$ matrices of the form
  \begin{equation*}
    \begin{pmatrix}
      0 & A \\ A^* & 0
    \end{pmatrix}~,
  \end{equation*}
  where $A$ is a complex $m\times n$ matrix and its hermitian adjoint $A^*$ is therefore a complex $n \times m$ matrix.  \emph{Mutatis mutandis} the
  same construction as above gives rise to a generalised metric Lie 3-algebra to which one may associate the pair $(\fsu(m) \oplus \fsu(n) \oplus
  \fu(1), \rf{(\boldsymbol{m},\overline{\boldsymbol{n}})})$.  The inner product on $\fsu(m) \oplus \fsu(n) \oplus \fu(1)$ is again indefinite, given by
  $\left(X_L\oplus X_R, Y_L\oplus Y_R\right) = \tr (X_L Y_L - X_RY_R)$, for $X_L,Y_L \in \fsu(m)$ and $X_R,Y_R \in \fsu(n)$, whereas the inner product
  on the $\fu(1)$ factor is either positive-definite or negative-definite depending on whether $m<n$ or $m>n$, respectively.  Of course, if $m=n$ we are
  back in the original case, in which the $\fu(1)$ factor is absent since it acts trivially on the space of matrices.
\end{example}

Finally, a further explicit example of generalised metric Lie 3-algebra appears in \cite{Nambu:1973qe} and is built out of the octonions.  It was shown
in \cite{Yamazaki:2008gg} that it satisfies the axioms (CS1)-(CS3).  This example may be deconstructed into the pair $(\fg_2, \OO)$, where $\OO$ denotes
the octonions, which is a reducible, faithful, orthogonal representation of $\fg_2$.  The resulting 3-algebra has a nondegenerate centre spanned by
$1\in\OO$.  Quotienting by the centre gives another generalised metric Lie 3-algebra associated with the pair $(\fg_2, \Im\OO)$, with $\Im\OO$ the
7-dimensional representation of imaginary octonions.

\subsection{Reconstruction}
\label{sec:reconstruction-CS}

Now let $\fg$ be a metric real Lie algebra and let $V$ be a faithful real orthogonal representation of $\fg$.  We will let $\left(-,-\right)$ and
$\left<-,-\right>$ denote the inner products on $\fg$ and $V$, respectively.  They are both invariant under the corresponding actions of $\fg$.

We will define a 3-bracket on $V$ as follows.  We start by defining a bilinear map $D: V \times V \to \fg$, by transposing the $\fg$-action.
Indeed, given $x,y\in V$, let $D(x,y)\in \fg$ be defined by
\begin{equation}
  \label{eq:D-CS}
  \left(X, D(x,y)\right) = \left<X \cdot x, y \right>\quad\forall X\in \fg~,
\end{equation}
where $\cdot$ stands for the action of $\fg$ on $V$.

\begin{remark}\label{re:KSA}
  This construction is reminiscent of the definition of the Dirac current associated to two spinors, which plays such an important rôle in the
  construction of the Killing superalgebra of a supersymmetric supergravity background.  (See, e.g., \cite{FMPHom}.)  In that case, the rôle of $V$ is
  played by spinors, whereas that of $\fg$ by vectors, with $\cdot$ being the Clifford action.
\end{remark}

\begin{lemma}\label{le:D-onto-CS}
  $D$ is surjective onto $\fg$.
\end{lemma}

\begin{proof}
  We will first show that the image of $D$ is an ideal.  We do this because this result is slightly more general, as it does not use the faithfulness of
  the action of $\fg$ on $V$, and results in a useful explicit formula.  The fact that the image of $D$ is an ideal follows by an explicit computation.
  Let $X\in \fg$ and $v,w\in V$.  Then for all $Y \in \fg$ we have
  \begin{align*}
    \left([D(v,w),X],Y\right) &= \left(D(v,w),[X,Y]\right) &&\text{since $\fg$ is metric}\\
    &= \left<[X,Y]\cdot v, w\right> &&\text{by \eqref{eq:D-CS}}\\
    &= \left<X \cdot Y \cdot v, w\right> - \left<Y \cdot X \cdot v, w\right> && \text{since $V$ is a $\fg$-module}\\
    &= - \left<Y \cdot v, X \cdot w\right> - \left<Y \cdot X \cdot v, w\right> && \text{since $X\in\fso(V)$}\\
    &= - \left(D(v,X \cdot w), Y \right) - \left(D(X\cdot v, w),Y\right) && \text{again by \eqref{eq:D-CS}}~,
  \end{align*}
  whence abstracting $Y$,
  \begin{equation}
    \label{eq:ideal-CS}
    [X,D(v,w)] = D(X\cdot v, w) + D(v, X\cdot w)~.
  \end{equation}
  Of course, this just says that $D$ is $\fg$-equivariant.

  Now let $X \in \fg$ be perpendicular to the image of $D$.  This means that $\left(X,D(v,w)\right) = 0$ for all $v,w\in V$, or equivalently that
  $\left<X\cdot v, w\right> = 0$ for all $v,w\in V$.  Nondegeneracy of the inner product on $V$ implies that $X\cdot v = 0$ for all $v\in V$, which in
  turn implies that $X=0$ since the action of $\fg$ on $V$ is faithful.  Finally, nondegeneracy of the inner product on $\fg$ says that $(\im D)^\perp =
  0$ implies surjectivity.
\end{proof}

Now let us define a trilinear map $V\times V \times V \to V$ by $(x,y,z)\mapsto [x,y,z] := D(x,y) \cdot z$.  We claim that this trilinear map turns $V$
into a generalised metric Lie 3-algebra.

\begin{proposition}
  The 3-bracket $[x,y,z] = D(x,y) \cdot z$ on $V$ obeys the axioms \eqref{eq:unitarity-CS}, \eqref{eq:symmetry-CS} and \eqref{eq:FI-CS} of
  Definition~\ref{def:CS}.
\end{proposition}

\begin{proof}
  The first axiom \eqref{eq:unitarity-CS} follows from the fact that $\fg$ preserves the inner product on $V$.

  The second axiom is the symmetry of the inner product on $\fg$.  Indeed, according to \eqref{eq:D-CS},
  \begin{equation*}
    \left(D(x,y), D(z,w)\right) = \left<D(x,y) \cdot z, w\right> = \left<[x,y,z],w\right>~,
  \end{equation*}
  whereas
  \begin{equation*}
    \left(D(z,w), D(x,y)\right) = \left<D(z,w) \cdot x, y\right> = \left<[z,w,x],y\right>~.
  \end{equation*}
  The equality of both of these expressions is precisely axiom \eqref{eq:symmetry-CS}.

  Finally, we prove the fundamental identity \eqref{eq:FI-CS} by substituting $X = D(x,y)$ into equation \eqref{eq:ideal-CS} in the proof of
  Lemma~\ref{le:D-onto-CS} and applying both sides of the equation to $z$.
\end{proof}

An \textbf{isomorphism} $\varphi: V \to W$ of generalised metric Lie 3-algebras is a linear map such that for all $x,y,z\in V$,
\begin{equation*}
  [\varphi x, \varphi y, \varphi z]_W = \varphi[x,y,z]_V \qquad\text{and}\qquad \left<\varphi x, \varphi y\right>_W = \left<x,y\right>_V~.
\end{equation*}
This last condition induces a Lie algebra isomorphism $\fso(V) \to \fso(W)$ by $X \mapsto \varphi \circ X \circ \varphi^{-1}$ which relates the Lie
algebras $\fg_V$ and $\fg_W$ by $\fg_W = \varphi \circ \fg_V \circ \varphi^{-1}$.  We say that pairs $(\fg_V,V)$ and $(\fg_W,W)$ are \textbf{isomorphic}
if there is an isometry $\varphi: V \to W$ which relates $\fg_V < \fso(V)$ and $\fg_W < \fso(W)$ by $\fg_W = \varphi \circ \fg_V \circ \varphi^{-1}$ and
such that this is an isometry of metric Lie algebras.  It then follows that isomorphic 3-algebras give rise to isomorphic pairs, and conversely.

It is clear that deconstruction and reconstruction are mutual inverse procedures, whence we have proved the following

\begin{theorem}
  There is a one-to-one correspondence between isomorphism classes of generalised metric Lie 3-algebras $V$ and isomorphic classes of pairs $(\fg, V)$
  where $\fg$ is a metric real Lie algebra and $V$ is a faithful orthogonal $\fg$-module.
\end{theorem}

This reduces the classification problem of generalised metric Lie 3-algebras to that of (conjugacy classes of) \emph{metric} Lie subalgebras $\fg <
\fso(V)$.  In other words, pairs $(\fg, b)$, where $\fg < \fso(V)$ and $b$ is an ad-invariant inner product on $\fg$.  It is important to remark that
the inner product on $\fg$ is an essential part of the data and that, in fact, it need not come induced from an inner product on $\fso(V)$.  To
illustrate this, consider the following example.

\begin{example}\label{eg:metric}
  Consider the pair $(\fso(4), \RR^4)$ of Example \ref{eg:A4}, but where the inner product on $\fso(4)$ is given by the Killing form, so that it is
  negative-definite instead of split.  It is easy to see that relative to an orthogonal basis $(\be_1,\dots,\be_4)$ for $\RR^4$, the 3-bracket is now
  given by
  \begin{equation*}
    [\be_i, \be_j, \be_k] = \delta_{jk} \be_i - \delta_{ik} \be_j~,
  \end{equation*}
  which, since $[x,y,z] + \textrm{cyclic} = 0$, defines on $\RR^4$ the structure of a Lie triple system.
\end{example}

\subsection{Recognition}
\label{sec:recognition-CS}

We saw above that there are two extreme cases of generalised metric Lie 3-algebras: the metric Lie 3-algebras of \cite{Filippov,FOPPluecker} and the
(metric) Lie triple systems \cite{JacobsonJordan,Lister,LTSreducedaxioms}.

It is possible to characterise the Lie triple systems among these 3-algebras as those for which on $\fg \oplus V$ one can define the structure of a
$\ZZ_2$-graded metric Lie algebra as follows.  We declare $\fg$ to be the even subspace and $V$ to be the odd subspace, with the following Lie
brackets extending those of $\fg$:
\begin{equation*}
  [D(x,y), z] = [x,y,z] \qquad\text{and}\qquad [x,y] = D(x,y)~.
\end{equation*}
Notice that since $D(x,y) = - D(y,x)$, the odd-odd bracket is skewsymmetric and hence this construction leads to a Lie algebra and not a Lie
superalgebra.  Being $\ZZ_2$-graded, there are four components to the Jacobi identity:
\begin{itemize}
\item the even-even-even (or $000$) component is the Jacobi identity for $\fg$;
\item the even-even-odd (or $001$) component is simply the fact that $V$ is a $\fg$-module;
\item the even-odd-odd (or $011$) component is the $\fg$-invariance of $D$, which is the content of equation \eqref{eq:ideal-CS}; and
\item the odd-odd-odd (or $111$) component is
  \begin{equation*}
    [[u,v],w] + \text{cyclic} = [D(u,v),w] + \text{cyclic} = [u,v,w] + \text{cyclic} = 0~,
  \end{equation*}
  which singles out the case where $V$ is a Lie triple system with $\fg\oplus V$ its embedding Lie algebra, as explained for example in \cite{Lister}.
\end{itemize}
We define an inner product on $\fg \oplus V$ by using the $\fg$-invariant inner products on $\fg$ and $V$, and declaring that $\fg \perp V$.  One checks
that this inner product on $\fg \oplus V$ is ad-invariant, turning $\fg \oplus V$ into a metric Lie algebra.

\begin{example}\label{eq:metric2}
  For the Lie triple system of Example~\ref{eg:metric}, the embedding Lie algebra is isomorphic to $\fso(5)$, whence the Lie triple system corresponds
  to the symmetric space $\SO(5)/\SO(4)$, which is the round 4-sphere.
\end{example}

\section{The hermitian 3-algebras of Bagger--Lambert}
\label{sec:BL4}

In \cite{BL4}, Bagger and Lambert constructed the three-dimensional $N{=}6$ superconformal field theory of \cite{MaldacenaBL} from a hermitian vector
space $(V,h)$ with a 3-bracket $[x,y;z]$ defined as follows.

Let $(V,h)$ be a finite-dimensional complex hermitian vector space, where $h$ is a nondegenerate sesquilinear map $h: V \times V \to \CC$ obeying
$h(v,w) = \overline{h(w,v)}$ and $h(\zeta v,w) = \zeta h(v,w)$, hence $h(v, \zeta w) = \overline{\zeta} h(v,w)$, for all $v,w\in V$ and $\zeta \in \CC$.
A hermitian structure defines a complex antilinear map $V \to V^*$ by $w \mapsto \hfl{w} := h(-,w)$, which is nevertheless an isomorphism of the
underlying real vector spaces.  The 3-bracket $[x,y;z]$ is complex linear in the first two entries and complex antilinear in the third and obeys the
following properties for all $u,v,w,x,y,z\in V$:
\begin{enumerate}
\item[(BL1)] the \emph{skewsymmetry} axiom:
  \begin{equation}
    \label{eq:skew-BL4}
    [x,y; z] = - [y,x; z]~;
  \end{equation}
\item[(BL2)] the \emph{symmetry} axiom
  \begin{equation}
    \label{eq:symmetry-BL4}
    h([x,y; z], w) = h(y, [z,w; x])~;
  \end{equation}
\item[(BL3)] and a \emph{fundamental identity}
  \begin{equation}
    \label{eq:FI-BL4}
    [x, [z,u;w];v] - [x,[z,u;v]; w] + [x,z; [w,v;u]] - [x,u;[w,v;z]] = 0~.
  \end{equation}
\end{enumerate}
\emph{Mutatis mutandis}, axioms \eqref{eq:skew-BL4} and \eqref{eq:symmetry-BL4} are precisely equation (35) in \cite{BL4}, whereas the
fundamental identity \eqref{eq:FI-BL4} is precisely equation (36) in \cite{BL4}.  Notice that axioms \eqref{eq:skew-BL4} and \eqref{eq:symmetry-BL4}
together imply
\begin{equation}
  \label{eq:skewtoo}
  h([x,y; z], w)  =  - h([x,y; w], z)~.
\end{equation}

We will find it convenient to rewrite the fundamental identity.

\begin{lemma}\label{le:equiv-FI}
  The fundamental identity \eqref{eq:FI-BL4} is equivalent to
  \begin{equation}
    \label{eq:FI-C}
    [[z,v;w],x;y] - [[z,x;y],v;w] - [z,[v,x;y];w] + [z,v;[w,y;x]] = 0~.
  \end{equation}
\end{lemma}

\begin{proof}
  First we take the hermitian inner product of equation \eqref{eq:FI-C} with $u$ to obtain
  \begin{equation*}
    h([[z,v;w],x;y],u) - h([[z,x;y],v;w],u) - h([z,[v,x;y];w],u) + h([z,v;[w,y;x]],u) = 0~,
  \end{equation*}
  and then bring this expression to a standard form without nested brackets.  To this end we use axiom \eqref{eq:skew-BL4} on the first two terms, axiom
  \eqref{eq:symmetry-BL4} on all but the last term of the resulting expression and then relation \eqref{eq:skewtoo} on the last term to obtain
  \begin{multline}
    \label{eq:FI-C-normal}
    h([z,v;w], [u,y;x]) - h([z,x;y], [u,w;v])\\ - h([v,x;y], [w,u;z]) + h([v,z;u],[w,y;x])=0~.
  \end{multline}
  We now bring equation \eqref{eq:FI-BL4} to a similar normal form, by taking the hermitian inner product with $y$ to obtain
  \begin{equation*}
    h([x, [z,u;w];v],y) - h([x,[z,u;v]; w],y) + h([x,z; [w,v;u]],y) - h([x,u;[w,v;z]],y) = 0~,
  \end{equation*}
  and use axiom \eqref{eq:symmetry-BL4} on the first two terms and \eqref{eq:skewtoo} on the last two terms to obtain
  \begin{equation*}
    h([z,u;w], [v,y;x]) - h([z,u;v], [w,y;x]) -  h([x,z; y], [w,v;u]) + h([x,u;y],[w,v;z]) = 0~.
  \end{equation*}
  Comparing with \eqref{eq:FI-C-normal}, we see that they are the same expression with $u,v$ interchanged after rearranging and using
  \eqref{eq:skew-BL4}.
\end{proof}

Just as we did for the generalised metric Lie 3-algebras of Section~\ref{sec:CS}, we will associate a pair $(\fg,V)$ consisting of a metric real Lie
algebra $\fg$ of which $V$ is a complex unitary representation.  In trying to reconstruct the 3-algebra from this data, we will actually arrive at a
more general class of hermitian 3-algebras.  The hermitian 3-algebras of \cite{BL4} will arise out of $(\fg,V)$ precisely when $\fg \oplus \rf{V}$
admits the structure of a Lie superalgebra into whose complexification $V$ embeds in a manner similar to the case of Lie triple systems in
Section~\ref{sec:recognition-CS}.

\subsection{Deconstruction}
\label{sec:deconstruction-BL4}

We start by defining a sesquilinear map $D: V \times V \to \End V$ from the 3-bracket by $D(v,w) z := [z,v;w]$.

\begin{lemma}
  The image of $D$ is a Lie subalgebra of $\fgl(V)$.
\end{lemma}

\begin{proof}
  By definition,
  \begin{align*}
    [ D(x,y), D(v,w) ] z &= D(x,y) [z,v;w] - D(v,w) [z,x;y]\\
    &= [[z,v;w],x;y] - [[z,x;y],v;w]\\
    &= [z,[v,x;y];w] - [z,v;[w,y;x]] && \text{by \eqref{eq:FI-C}.}
  \end{align*}
  In terms of $D$, this equation becomes
  \begin{equation}
    \label{eq:DD-BL4}
    [ D(x,y), D(v,w) ] z = D([v,x;y],w) z - D(v,[w,y;x])z~.
  \end{equation}
  Finally, abstracting $z$ we see that the image of $D$ closes under the commutator and is hence a Lie subalgebra of $\fgl(V)$.
\end{proof}

We let $\fg$ denote the span of the $D(x,y)$ in $\End V$.  It is a complex Lie algebra.  Conditions \eqref{eq:skew-BL4} and \eqref{eq:symmetry-BL4}
together imply that
\begin{equation*}
  h(D(x,z)y , w) - h(y, D(z,x)w) = 0~,
\end{equation*}
which we would like to interpret as a unitarity condition.  We have to be careful, though, since $\fg$ is a complex Lie algebra and hence cannot
preserve a hermitian inner product.  The notion of unitarity for complex Lie algebras says that
\begin{equation*}
  h(Xy , w) + h(y, c(X)w) = 0~,
\end{equation*}
where $c$ is a \textbf{conjugation} on $\fg$; that is, $c$ is a complex antilinear, involutive automorphism of $\fg$.  These conditions on
$c$ guarantee that its fixed subspace
\begin{equation*}
  \fg^c := \left\{X \in \fg \middle | c(X) = X\right\} < \fu(V)
\end{equation*}
is a real Lie algebra, said to be a \textbf{real form} of $\fg$, which then does leave $h$ invariant.

This suggests defining $c: \fg \to \fg$ by $c D(x,y) = - D(y,x)$.  From $D( i x, y) = i D(x,y)$ and $D(x, i y) = -i D(x,y)$ for all $x,y\in V$, it
follows that $c(D(i x, y)) = c(i D(x,y))$, whereas $-D(y,i x) = i D(y,x) = -i c(D(x,y))$, which shows that $c$ can be extended to a complex
antilinear map to all of $\fg$.  It is clear moreover that $c$ is involutive.  Finally, it remains to show that $c$ is an automorphism.  To do this we
compute
\begin{align*}
  [c D(x,y),c D(u,v)] &= [-D(y,x), -D(v,u)]\\
  &= [D(y,x), D(v,u)]\\
  &= D(D(y,x)v, u) - D(v, D(x,y)u)\\
  &= c( D(D(x,y)u, v) - D(u, D(y,x)v))\\
  &= c[D(x,y), D(u,v)]~.
\end{align*}

The real form $\fg^c$ is then spanned by $E(x,y):= D(x,y) - D(y,x)$ for all $x,y\in V$.  For example, if $(\be_a)$ is a complex basis for $V$, then
$\fg^c$ is spanned by $D(\be_a, \be_b)- D(\be_b,be_a)$ and $i(D(\be_a,\be_b) + D(\be_b,\be_a))$.  We will now show that $\fg^c$ is a metric
Lie algebra.

\begin{proposition}\label{pr:IPonG}
  The bilinear form on $\fg^c$ defined by
  \begin{equation*}
    \left(E(x,y), E(u,v)\right) := \Re h(E(x,y)u, v)~.
  \end{equation*}
  is symmetric, nondegenerate and ad-invariant.
\end{proposition}

\begin{proof}
  We prove each property in turn.  To prove symmetry we simply calculate:
  \begin{align*}
    h(E(x,y)u, v) &= h(D(x,y)u, v)- h(D(y,x)u, v)\\
    &= h([u,x;y], v) - h([u,y;x], v)\\
    &= h(x,[y,v;u]) - h(y,[x,v;u]) &&\text{using \eqref{eq:symmetry-BL4}}\\
    &= h(x,D(v,u)y) - h(y, D(v,u)x)\\
    &= h(D(u,v)x, y) - \overline{h(D(v,u)x, y)}~,
  \end{align*}
  whence taking real parts we find
  \begin{equation*}
    \Re h(E(x,y)u, v) = \Re h (E(u,v) x, y)~.
  \end{equation*}

  To prove nondegeneracy, let us assume that some linear combination $X = \sum_i E(x_i, y_i)$ is orthogonal to all $E(u,v)$ for $u,v\in V$:
  \begin{equation*}
    \Re h(X u, v) = 0~.
  \end{equation*}
  Now, since $h$ is nondegenerate, so is $\Re h$ because $\Im h(x,y) = \Re h(-i x, y)$, whence this means that $X u = 0$ for all $u$, showing
  that the endomorphism $X =0$.

  Finally, we show that it is ad-invariant.   A simple calculation using equation \eqref{eq:DD-BL4} shows that
  \begin{equation*}
    [E(x,y), E(u,v)] = E(E(x,y)u, v) + E(u, E(x,y)v)~,
  \end{equation*}
  whence
  \begin{align*}
    \left(E(z,w),[E(x,y),E(u,v)]\right) &= \left(E(z,w), E(E(x,y)u, v) + E(u, E(x,y)v)\right)\\
    &= \Re h(E(z,w)E(x,y) u, v) + \Re h( E(z,w)u, E(x,y) v)\\
    &= \Re h(E(z,w)E(x,y) u, v) - \Re h( E(x,y) E(z,w)u, v)\\
    &= \Re h([E(z,w), E(x,y)] u, v) \\
    &= \left([E(z,w), E(x,y)], E(u,v)\right)~.
  \end{align*}
\end{proof}

In summary, we have extracted a metric real Lie algebra $\fg^c$ and a unitary representation $(V,h)$ from the hermitian 3-algebra in \cite{BL4}.  The
Lie algebra $\fg^c$ is the gauge algebra in the $N{=}6$ theory.  We shall illustrate this construction with the explicit example in \cite{BL4}, which is
very similar in spirit to Example \ref{eg:CS}.

\begin{example}\label{eg:BL4}
  Let $V$ denote the space of $m \times n$ complex matrices.  For all $x,y,z \in V$, define
  \begin{equation*}
    [x,y;z] := y z^* x - x z^* y~,
  \end{equation*}
  where $z^*$ denotes the hermitian adjoint of $z$.  The generators of $\fg^c$ are $E(x,y) = D(x,y) - D(y,x)$, where $D(y,z)x = [x,y;z]$.
  The action of $E(x,y)$ on $V$ is given by
  \begin{align*}
    E(y,z) x &= D(y,z) x - D(z,y) x\\
    &= y z^* x - x z^* y - z y^* x + x y^* z \\
    &= x (y^* z - z^* y) + (y z^* - z y^*) x
  \end{align*}
  whence it consists of a linear combination of left multiplication by the skewhermitian $m\times m$ matrix $y z^*-z y^*$ and right multiplication by
  the skewhermitian $n\times n$ matrix $y^* z-z^* y$.  Let us decompose them into traceless + scalar matrices as follows
  \begin{equation*}
    y z^* - z y^* = A(y,z) + \tfrac{i}{m}\alpha(y,z) 1_m \qquad\text{and}\qquad
    y^* z - z^* y = B(y,z) - \tfrac{i}{n}\alpha(y,z) 1_n~, 
  \end{equation*}
  where $A(y,z)$ and $B(y,z)$ are traceless, whence in $\fsu(m)$ and $\fsu(n)$, respectively, and $\alpha(y,z) = 2 \Im\tr y z^*$.  Into the action of
  $E(y,z)$ on $x$, we find
  \begin{equation*}
    E(y,z) x = A(y,z) x + x B(y,z) + i \alpha(y,z) (\tfrac1m - \tfrac1n) x~.
  \end{equation*}
  The hermitian inner product on $V$ is given by $h(x,y) = 2 \tr x y^*$, where the factor of $2$ is for later convenience.  The invariant inner product
  on $\fg^c$ in Proposition~\ref{pr:IPonG} is given by polarizing
  \begin{equation*}
    \left(E(x,y),E(x,y)\right) = \tr A(x,y)^2 - \tr B(x,y)^2  - \alpha(x,y)^2  (\tfrac1m - \tfrac1n)~.
  \end{equation*}
  Therefore we see that if $m=n$, then $\fg^c \cong \fsu(n) \oplus \fsu(n)$ acting on $V \cong \CC^n \otimes (\CC^n)^*$, which is the bifundamental
  $(\boldsymbol{n},\overline{\boldsymbol{n}})$, and the inner product is given by $\left(X_L\oplus X_R, Y_L\oplus Y_R\right) = \tr (X_L Y_L) - \tr (X_R
  Y_R)$, for $X_L,Y_L,X_R,Y_R\in\fsu(n)$, with the traces in the fundamental.  If $m\neq n$, then $\fg^c \cong \fsu(m) \oplus \fsu(n) \oplus \fu(1)$,
  which is the quotient of $\fu(m) \oplus \fu(n)$ by the kernel of its action on $V$.  The inner product on the semisimple part is the same as in the
  case $m = n$, whereas the inner product on the centre is positive-definite (respectively, negative-definite) accordingly to whether $m<n$
  (respectively, $m>n$).
\end{example}

\subsection{Reconstruction}
\label{sec:reconstruction-BL4}

We would like to reconstruct a hermitian 3-algebra out of a pair $(\fk,V)$ where $\fk$ is a (real) metric Lie algebra and $(V,h)$ is a faithful
(complex) unitary representation, as we did for the generalised metric Lie 3-algebras in Section \ref{sec:reconstruction-CS}.  In the attempt, we will
find that we obtain a more general class of algebras than the hermitian 3-algebras of \cite{BL4}.  This prompts us to understand the conditions on
$(\fk,V)$ which will guarantee that we do obtain the class of hermitian 3-algebras of \cite{BL4}.  We will see in Section \ref{sec:recognition-BL4} that
this condition says that $\fk \oplus \rf{V}$ admits the structure of a Lie superalgebra satisfying some natural properties.

Let $\fk$ be a real metric Lie algebra with invariant inner product $\left(-,-\right)$ and let $(V,h)$ be a complex hermitian vector space admitting a
faithful unitary action of $\fk$:
\begin{equation}
  \label{eq:unitaryrep}
  h(X \cdot v, w) + h(v,X \cdot w) = 0 \qquad\text{for all $X\in\fk$, $v,w\in V$.}
\end{equation}
Let $\fk_\CC = \CC \otimes_{\RR}\fk$ denote the complexification of $\fk$, turned into a complex Lie algebra by extending the Lie bracket on $\fk$
complex bilinearly.  We also extend the inner product complex bilinearly.  As it remains nondegenerate, it turns $\fk_\CC$ into a complex metric Lie
algebra.  Furthermore we extend the action of $\fk$ on $V$ to $\fk_\CC$ by $(X+ i Y)\cdot v = X\cdot v + i Y\cdot v$, for all $X,Y \in \fk$ and $v\in
V$.  As mentioned above, a complex Lie algebra cannot leave a hermitian inner product invariant.  Instead we have that
\begin{equation}
  \label{eq:pre-unitarity-C}
  h(\XX \cdot v , w) + h(v, \XXbar\cdot w) = 0~,\qquad\text{for all $\XX \in \fk_\CC$ and $v,w\in V$.}
\end{equation}
Indeed, letting $\XX = X + i Y$,
\begin{align*}
  h(\XX\cdot v, w) &= h((X + iY) \cdot v, w)\\
  &= h(X\cdot v, w) + i h(Y\cdot v, w)\\
  &= - h(v, X \cdot w) - i h(v, Y \cdot w) && \text{by equation~\eqref{eq:unitaryrep}}\\
  &= - h(v, X \cdot w) + h(v, i Y \cdot w) && \text{by sesquilinearity of $h$}\\
  &= - h(v, (X-iY)\cdot w)\\
  &= - h(v,\XXbar\cdot w)~.
\end{align*}

\begin{lemma}
  $V$ remains a faithful representation of $\fk_\CC$.
\end{lemma}

\begin{proof}
  Let $\XX = X + i Y \in \fk_\CC$ be such that $\XX \cdot v = 0$ for all $v\in V$.  We want to show that $\XX = 0$.  Taking the complex conjugate of
  that equation says that $\XXbar \cdot v = 0$ for all $v\in V$.  Therefore the real and imaginary parts of $\XX$ satisfy the same equation: $X \cdot v
  = 0$ and $Y \cdot v = 0$ for all $v\in V$.  Since $X,Y\in \fk$ and $\fk$ acts faithfully on $V$, $X=Y=0$ and hence $\XX = 0$.
\end{proof}

Let us define a sesquilinear map $D: V \times V \to \fk_\CC$ as follows.  Let $x,y\in V$ and define $D(x,y)$ by
\begin{equation}
  \label{eq:D-C}
  \left( D(x,y), \XX \right) = h(\XX\cdot x, y)~,
\end{equation}
for all $\XX \in \fk_\CC$.

\begin{lemma}\label{le:D-onto-BL4}
  $D$ is surjective onto $\fk_\CC$.
\end{lemma}

\begin{proof}
  The proof follows along the same lines as that of Lemma~\ref{le:D-onto-CS}.  We prove that the image of $D$ is an ideal by explicit computation.
  Let $\XX\in \fk_\CC$ and $v,w\in V$.  Then for all $\YY \in \fk_\CC$ we have
  \begin{align*}
    \left([D(v,w),\XX],\YY\right) &= \left(D(v,w),[\XX,\YY]\right) &&\text{since $\fk_\CC$ is metric}\\
    &= h([\XX,\YY]\cdot v, w) &&\text{by \eqref{eq:D-C}}\\
    &= h(\XX \cdot \YY \cdot v, w) - h(\YY \cdot \XX \cdot v, w) && \text{since $V$ is a $\fk_\CC$-module}\\
    &= - h(\YY \cdot v, \XXbar \cdot w) - h(\YY \cdot \XX \cdot v, w) && \text{by \eqref{eq:pre-unitarity-C}}\\
    &= - \left(D(v,\XXbar \cdot w), \YY \right) - \left(D(\XX\cdot v, w),\YY\right) && \text{again by \eqref{eq:D-C}}~,
  \end{align*}
  whence abstracting $\YY$,
  \begin{equation}
    \label{eq:ideal-C}
    [\XX,D(v,w)] = D(\XX\cdot v, w) + D(v, \XXbar\cdot w)~.
  \end{equation}

  Now let $\XX \in \fk_\CC$ be perpendicular to the image of $D$.  This means that $\left(\XX,D(v,w)\right) = 0$ for all $v,w\in V$, or equivalently
  that $h(\XX\cdot v, w) = 0$ for all $v,w\in V$.   Nondegeneracy of $h$ implies that $\XX\cdot v = 0$ for all $v\in V$, which in turn
  implies that $\XX=0$ since the action of $\fk_\CC$ on $V$ is still faithful.  Nondegeneracy of the inner product on $\fk_\CC$ says that $(\im D)^\perp
  = 0$ implies surjectivity.
\end{proof}

We now define a ternary product: $V \times V \times V \to V$, complex linear in the first and third entries and complex antilinear in the second, by
\begin{equation}
  \label{eq:bracket-C}
  (x,y,z) \mapsto [z,x;y] := D(x,y)\cdot z~,
\end{equation}
where the notation has been chosen to ease the comparison with the 3-bracket in \cite{BL4}.

\begin{lemma}
  In terms of the 3-bracket, the unitarity condition \eqref{eq:pre-unitarity-C} reads
  \begin{equation}
    \label{eq:unitarity-C}
    h([x,v;w],y) = h(x,[y,w;v])~.
  \end{equation}
\end{lemma}

\begin{proof}
  From equation \eqref{eq:pre-unitarity-C}, we have that
  \begin{equation}
    \label{eq:pre-unitarity-C-too}
    h(D(v,w)\cdot x, y) = - h(x,  \overline{D(v,w)}\cdot y)~.
  \end{equation}
  Complex conjugating the definition \eqref{eq:D-C} of $D(v,w)$, we find on the one hand
  \begin{equation*}
    \overline{\left(D(v, w), \XX\right)} =  \left(\overline{D(v,w)}, \XXbar\right)~,
  \end{equation*}
  and on the other
  \begin{align*}
    \overline{\left(D(v,w), \XX\right)} &= \overline{h(\XX \cdot v,w)}\\
    &= h(w,\XX \cdot v)\\
    &= - h(\XXbar \cdot w, v)  && \text{using \eqref{eq:pre-unitarity-C}}\\
    &= - \left(D(w,v), \XXbar\right)~,
  \end{align*}
  whence
  \begin{equation}
    \label{eq:Dbar}
    \overline{D(v,w)} = - D(w,v)~.
  \end{equation}
  Substituting this into equation \eqref{eq:pre-unitarity-C-too},
  \begin{equation*}
    h(D(v,w)\cdot x, y) = h(x,  D(w,v)\cdot y)~,
  \end{equation*}
  which is precisely equation \eqref{eq:unitarity-C} when written in terms of the 3-bracket.
\end{proof}

The symmetry of the complex inner product on $\fk_\CC$ is
\begin{equation*}
  \left(D(x,y),D(v,w)\right) = \left(D(v,w),D(x,y)\right)~,
\end{equation*}
which becomes
\begin{equation*}
  h(D(v,w)\cdot x, y) = h(D(x,y)\cdot v, w)~,
\end{equation*}
or equivalently
\begin{equation}
  \label{eq:symmetry-C}
  h([x,v;w],y) = h([v,x;y],w)
\end{equation}
in terms of the 3-bracket.

Substituting $\XX = D(x,y)$ in equation \eqref{eq:ideal-C} and using equation~\eqref{eq:Dbar}, leads to the fundamental identity
\begin{equation}
  \label{eq:[D,D]-C}
  [D(x,y),D(v,w)] = D(D(x,y)\cdot v,w) - D(v, D(y,x)\cdot w)~.
\end{equation}
Applying this equation to $z$ and writing everything in terms of the 3-bracket, we arrive at the fundamental identity in \eqref{eq:FI-C}.

In summary, the most general 3-algebra constructed from a metric Lie algebra and a faithful complex unitary representation $V$ is defined by a 3-bracket
$[x,y;z]$, complex linear in the first two entries and complex antilinear in the third, satisfying equations \eqref{eq:unitarity-C} and
\eqref{eq:symmetry-C}, which are linear in the 3-bracket, and the fundamental identity \eqref{eq:FI-C}, which is quadratic.  The class of such
3-algebras is strictly a superset of the hermitian 3-algebras in \cite{BL4}.  Indeed, it is easy to see that the antisymmetry condition
\eqref{eq:skew-BL4} does not follow from \eqref{eq:unitarity-C} and \eqref{eq:symmetry-C}.  Indeed, starting from $h([x,y;z],w)$ and applying
\eqref{eq:unitarity-C} and \eqref{eq:symmetry-C} in turn, one arrives at the following symmetry properties:
\begin{equation*}
  h([x,y;z],w) \stackrel{\eqref{eq:symmetry-C}}{=} h([y,x;w],z) \stackrel{\eqref{eq:unitarity-C}}{=} h(y,[z,w;x]) \stackrel{\eqref{eq:symmetry-C}}{=}
  h(x,[w,z;y])
\end{equation*}
and no others.  In particular, as there are no signs involved, one can never derive a skewsymmetry condition such as \eqref{eq:skew-BL4}.

Nevertheless we can easily characterise the hermitian 3-algebras in \cite{BL4} in terms of Lie superalgebras, in agreement with an observation in
\cite{SchnablTachikawa} based on \cite{3Lee,GaiottoWitten}.  This can be thought of as a purely algebraic explanation of this fact.

\subsection{Recognition}
\label{sec:recognition-BL4}

Consider the hermitian 3-algebra obtained above from a pair $(\fk,V)$ consisting of a (real) metric Lie algebra and a (complex) unitary faithful
representation $(V,h)$.  Recall that it satisfies the axioms \eqref{eq:unitarity-C}, \eqref{eq:symmetry-C} and \eqref{eq:FI-C}.  We attempt to construct
a Lie (super)algebra on the $\ZZ_2$-graded vector space $\fh = \fk_\CC \oplus (V \oplus V^*)$, where the even subspace is the span of the $D(x,y)$ and
the odd subspace is $V \oplus V^*$.  The nonzero brackets are given by the following natural maps
\begin{equation}\label{eq:superalgebra}
  \begin{aligned}[m]
    [D(x,y),z] &= D(x,y) \cdot z = [z,x;y]\\
    [D(x,y),\hfl{z}] &= - \hfl{D(y,x)\cdot z} = - \hfl{[z,y;x]}\\
    [x,\hfl{y}] &= D(x,y)~,
  \end{aligned}
\end{equation}
where we choose to parametrise $V^*$ by elements of $V$: $z \mapsto \hfl{z} = h(-,z)$.  As before, there are four components to the Jacobi identity.
The $000$-component is satisfied by virtue of the fact that $\fk_\CC$ is a Lie algebra, as shown by the fundamental identity \eqref{eq:FI-C}.
Similarly, the $001$-component is satisfied because $V\oplus V^*$ is a $\fk_\CC$-module.  Since the map $D$ is $\fk_\CC$-equivariant (again a
consequence of the fundamental identity), the $011$-component is also satisfied.  It remains, as usual, to check the $111$-component.  There are two
such Jacobi identities:
\begin{align*}
  [x,[y,\hfl{z}]] &= [[x,y],\hfl{z}] \pm [y,[x,\hfl{z}]]\\
  [\hfl{x},[\hfl{y},z]] &= [[\hfl{x},\hfl{y}],z] \pm [\hfl{y},[\hfl{x},z]]~,
\end{align*}
where the top sign is for the case of a $\ZZ_2$-graded Lie algebra and the bottom sign for a Lie superalgebra.  It is not hard to show that both
identities are equivalent, so we need only investigate the first, say.  Using that $[x,y]=0$ and the definition of the other brackets, we obtain
\begin{equation*}
  [x,[y,\hfl{z}]] = [x, D(y,z)] = - D(y,z)\cdot x = -[x,y;z]
\end{equation*}
and
\begin{equation*}
  \pm [y,[x,\hfl{z}]] = \pm [y, D(x,z)] = \mp D(x,z) \cdot y = \mp [y,x;z]~,
\end{equation*}
whence the condition is precisely
\begin{equation*}
  [x,y;z] = \pm [y,x;z]~,
\end{equation*}
where the top sign is for a $\ZZ_2$-graded Lie algebra, whereas the bottom is for a Lie superalgebra.  This latter case is precisely condition
\eqref{eq:skew-BL4} in the 3-algebra of \cite{BL4}, whence we conclude that these 3-algebras are characterised as those hermitian 3-algebras constructed
from $(\fk,V)$ for which $\fk_\CC \oplus V \oplus V^*$ becomes a (complex) Lie superalgebra.

\begin{remark}
  Alternatively, as discussed in \cite{NilssonPalmkvist}, one could try to embed the hermitian 3-algebra of \cite{BL4} in a graded Lie algebra, but even
  for the case of Example \ref{eg:BL4}, this graded Lie algebra is seemingly infinite-dimensional and not of Kac--Moody type.
\end{remark}

The complex Lie superalgebra $\fk_\CC \oplus V \oplus V^*$ is the complexification of a real Lie superalgebra $\fs = \fs_0 \oplus \fs_1$ defined as
follows.  Let $\fs_1$ denote the real subspace of the complex vector space $V\oplus V^*$ of the form
\begin{equation*}
  \fs_1 = \left\{x + i \hfl{x} \middle | x \in V\right\}~.
\end{equation*}
% This subspace is a rotation by $-\pi/4$ of the natural real subspace $\left\{ x + \hfl{x}\middle | x \in V\right\}$.
Computing the Lie bracket of two such vectors we find
\begin{align*}
  [x + i \hfl{x}, y + i \hfl{y}] &= i [x,\hfl{y}] + i [\hfl{x},y]\\
  &= i D(x,y) + i D(y,x)\\
  &= D(ix,y) - D(y,ix)\\
  &= E(ix,y)~.
\end{align*}
In other words, $[\fs_1,\fs_1] = \fk$, the real form of $\fk_\CC$ leaving the hermitian inner product invariant.  We therefore let $\fs_0 = \fk$.
Computing the Lie bracket of $\fk$ and $\fs_1$ we find
\begin{equation}
  \label{eq:k-preserves-s1}
  [E(x,y),z+ i \hfl{z}] = [E(x,y),z] + i \hfl{[E(x,y),z]} \in \fs_1~,
\end{equation}
whence $\fs_1$ is a $\fk$-module.  It is in fact isomorphic to the underlying real representation $\rf{V}$ of $V \oplus V^*$.  We conclude that $\fs$ is
a real form of the complex Lie superalgebra $\fk_\CC \oplus (V \oplus V^*)$ defined above.

Let us now show that $\fs$ is a metric Lie superalgebra.  The inner product is obtained by extending the inner product on $\fk$ given by (see
Proposition~\ref{pr:IPonG})
\begin{equation*}
  \left(E(x,y),E(u,v)\right) = \Re h(E(x,y) \cdot u , v)
\end{equation*}
to all of $\fs$ by ad-invariance.  Since the inner product has even parity, $\fs_0 \perp \fs_1$.  The inner product on $\fs_1$ is then defined as
follows.  Let us introduce the notation $\rf{z} = z + i \hfl{z}$, for $z\in V$.
\begin{align*}
  \left(\rf{[E(x,y),z]},\rf{w}\right) &=  \left([E(x,y), \rf{z}], \rf{w}\right) && \text{by \eqref{eq:k-preserves-s1}}\\
  &= \left(E(x,y), [\rf{z},\rf{w}]\right) && \text{by ad-invariance}\\
  &= \left(E(x,y), E(iz,w)\right) && \text{by the above calculation}\\
  &= \Re h([E(x,y),i z], w) &&\text{by definition}\\
  &= \Re i h([E(x,y),z], w) &&\text{by sesquilinearity of $h$}\\
  &= -\Im h([E(x,y),z], w)~,
\end{align*}
where $\Im h$ denotes the imaginary part of the hermitian structure, which is skewsymmetric.  Thus we define the inner product on $\fs_1$ to
be
\begin{equation*}
  \left(\rf{x},\rf{y}\right) := - \Im h (x,y)~,
\end{equation*}
which is symplectic, as we would expect for a `super-symmetric' bilinear form restricted to the odd subspace of a vector superspace.  The resulting
inner product on $\fs$ is ad-invariant by construction, making $\fs$ into a metric Lie superalgebra.

Conversely, we will show that a real metric Lie superalgebra $\fs = \fs_0 \oplus \fs_1$ satisfying the following additional properties:
\begin{enumerate}
\item $\fs_1 \cong \rf{V} = [V \oplus V^*]$, as an $\fs_0$-module, where $(V,h)$ is a complex unitary representation of $\fs_0$, and
\item the Lie brackets $[V,V]$ and $[V^*,V^*]$ vanish,
\end{enumerate}
defines a 3-algebra obeying the axioms in \cite{BL4}.  To see this, we will reconstruct the 3-bracket by first complexifying the Lie superalgebra to
$\fs_\CC = (\fs_0)_\CC \oplus (\fs_1)_\CC$, where $(\fs_1)_\CC = V \oplus V^*$, and extending the Lie bracket and the inner product complex-bilinearly.
This makes $\fs_\CC$ into a complex metric Lie superalgebra.

As before, we parametrise $V^*$ in terms of $V$, via the complex antilinear map $v \mapsto \hfl{v} = h(-,v)$.  This allows us to define a 3-bracket on
$V$ by nesting the Lie brackets $[-,-]$ of $\fs_\CC$ as follows:
\begin{equation*}
  [x,y;z] := [[y,\hfl{z}],x]~,
\end{equation*}
for all $x,y,z\in V$.  As the notation suggests, the 3-bracket is complex linear in the first entries and antilinear in the third.  We will now show
that this 3-bracket obeys conditions \eqref{eq:skew-BL4}, \eqref{eq:symmetry-BL4} and \eqref{eq:FI-BL4} at the start of this section.

Condition \eqref{eq:skew-BL4} is a simple consequence of the Jacobi identity:
\begin{equation*}
  [x,[y,\hfl{z}]] = [[x,y],\hfl{z}] - [y,[x,\hfl{z}]]~,\quad\text{for all $x,y,z\in V$},
\end{equation*}
using that $[x,y]=0$.  The left-hand side is $-[x,y;z]$ whereas the right-hand side is $[y,x;z]$.  Their equality is precisely \eqref{eq:skew-BL4}.

Condition \eqref{eq:symmetry-BL4} is a essentially a consequence of unitarity.  Let $\XX\in\fk_\CC$ and $v,w\in V$.  Then on the one hand
\begin{equation*}
  \left(\XX, [v,\hfl{w}]\right)  = \left([\XX,v],\hfl{w}\right) = h([\XX,v],w) = - h(v,[\XXbar, w])~,
\end{equation*}
and taking the complex conjugate
\begin{equation*}
 \left(\XXbar, \overline{[v,\hfl{w}]}\right) = - h([\XXbar, w],v) = - \left([\XXbar, w],\hfl{v}\right) = - \left(\XXbar, [w,\hfl{v}]\right)~,
\end{equation*}
whence
\begin{equation}
  \label{eq:barbracket}
  \overline{[v,\hfl{w}]}=-[w,\hfl{v}]~.
\end{equation}
Using this we now calculate
\begin{align*}
  h([x,y;z],w) &= - h([y,x;z],w) && \text{by \eqref{eq:skew-BL4}, already shown to hold}\\
  &= -h([[x,\hfl{z}],y],w)  &&\text{by definition of the 3-bracket}\\
  &= h(y, [\overline{[x,\hfl{z}]},w]) && \text{by ``unitarity"}\\
  &= -h(y, [[z,\hfl{x}],w]) &&\text{by \eqref{eq:barbracket}}\\
  &= -h(y, [w,z;x]) &&\text{by definition of the 3-bracket}\\
  &= h(y,[z,w;x]) &&\text{again by \eqref{eq:skew-BL4},}
\end{align*}
which is precisely condition \eqref{eq:symmetry-BL4}.

Finally, the fundamental identity \eqref{eq:FI-BL4} is equivalent, as shown in Lemma~\ref{le:equiv-FI}, to equation~\eqref{eq:FI-C}, which in turn is
equivalent to equation \eqref{eq:[D,D]-C}.  It is therefore sufficient to show that equation \eqref{eq:[D,D]-C} follows from the Jacobi identity of the
superalgebra defined in \eqref{eq:superalgebra}.  Starting with the left-hand side of equation \eqref{eq:[D,D]-C} and expanding, we obtain
\begin{align*}
  [D(x,y),D(v,w)] &= [[x,\hfl{y}],[v,\hfl{w}]] && \text{by \eqref{eq:superalgebra}}\\
  &= [[[x,\hfl{y}],v],\hfl{w}] + [v,[[x,\hfl{y}],\hfl{w}]] && \text{by Jacobi}\\
  &= [[[x,\hfl{y}],v],\hfl{w}] - [v,\hfl{[[y,\hfl{x}],w]}] && \text{by \eqref{eq:superalgebra}}\\
  &= D(D(x,y)\cdot v, w) - D(v,D(y,x)\cdot w) && \text{again by \eqref{eq:superalgebra}.}\\
\end{align*}

In summary, we have proved the following

\begin{theorem}\label{th:BL4}
  There is a one-to-one correspondence between the hermitian 3-algebras of \cite{BL4} and metric Lie superalgebras (into whose complexification one
  embeds the 3-algebra) satisfying that the odd subspace is $\rf{V} \cong [V \oplus V^*]$ for $(V,h)$ a complex unitary faithful representation of the
  even subalgebra, where $[V,V]=0=[V^*,V^*]$.
\end{theorem}

This is morally in agreement with a statement in \cite{SchnablTachikawa} based on \cite{3Lee,GaiottoWitten} and clarifies the relationship between these
works and \cite{BL4}.

In the explicit example of \cite{BL4}, corresponding to Example \ref{eg:BL4} here, this Lie superalgebra is a ``compact'' real form of the classical Lie
superalgebra $A(m-1,n-1)$.  Among the classical superalgebras, also $C(n+1)$ defines a hermitian 3-algebra of the type considered in \cite{BL4}.  For
the compact real form of $C(n+1)$, the even subalgebra is $\fusp(2n) \oplus \fu(1)$ and the odd subspace is the representation
$\rf{(\boldsymbol{2n},\boldsymbol{1})}$.

\section{A general algebraic construction}
\label{sec:generalisation}

In this section we show that the above reconstructions of 3-algebras are special cases of a general algebraic construction due to Faulkner
\cite{FaulknerIdeals}.

\subsection{The Faulkner construction}
\label{sec:Faulkner}

There are two ingredients in this construction: a finite-dimensional metric Lie algebra $\fg$ and a finite-dimensional representation $V$.  The Lie
algebra $\fg$, being metric, admits an ad-invariant nondegenerate symmetric bilinear form $\left(-,-\right)$.  We will assume that $\fg$ is either real
or complex and that so is the representation $V$.  The dual representation will be denoted $V^*$.  We will let $\left<-,-\right>$ denote the dual
pairing between $V$ and $V^*$.  This data allows us to define a linear map $\eD: V \otimes V^* \to \fg$ as follows.  If $v \in V$ and $\alpha \in V^*$,
then $\eD(v,\alpha)\in \fg$ is defined by
\begin{equation}
  \label{eq:D-map}
  \left(X,\eD(v,\alpha)\right) = \left<X \cdot v, \alpha\right> \quad\text{for all $X\in\fg$,}
\end{equation}
where the $\cdot$ indicates the action of $\fg$ on a module, here $V$.

\begin{lemma}\label{le:ImD-ideal}
  The image of $\eD$ is an ideal of $\fg$, which is all of $\fg$ if $V$ is a faithful representation.
\end{lemma}

\begin{proof}
  The fact that the image of $\eD$ is an ideal follows by an explicit computation.  Let $X\in \fg$, $v\in V$ and $\alpha \in V^*$.  Then for all $Y \in
  \fg$ we have
  \begin{align*}
    \left([\eD(v,\alpha),X],Y\right) &= \left(\eD(v,\alpha),[X,Y]\right) &&\text{since $\fg$ is metric}\\
    &= \left<[X,Y]\cdot v, \alpha\right> &&\text{by \eqref{eq:D-map}}\\
    &= \left<X \cdot Y \cdot v, \alpha\right> - \left<Y \cdot X \cdot v, \alpha\right>\\
    &= - \left<Y \cdot v, X \cdot \alpha\right> - \left<Y \cdot X \cdot v, \alpha\right> \\
    &= - \left(\eD(v,X \cdot \alpha), Y \right) - \left(\eD(X\cdot v, \alpha),Y\right) && \text{by \eqref{eq:D-map}}~,
  \end{align*}
  whence abstracting $Y$,
  \begin{equation}
    \label{eq:ideal}
    [X,\eD(v,\alpha)] = \eD(X\cdot v, \alpha) + \eD(v, X\cdot \alpha)~.
  \end{equation}
  Of course, this just says that $\eD$ is $\fg$-equivariant.

  Now let us assume that $V$ is a faithful representation and let $X\in\fg$ be perpendicular to $\im\eD$.  Then for all $v\in V$ and $\alpha\in V^*$,
  \begin{equation*}
    \left(X, \eD(v,\alpha)\right) = \left<X\cdot v, \alpha\right> = 0~.
  \end{equation*}
  This implies that $X \cdot v = 0$ for all $v\in V$, which implies in turn that $X=0$ because the representation of $\fg$ on $V$ is faithful.
  Therefore $(\im\eD)^\perp = 0$, whence $\eD$ is surjective.
\end{proof}

The map $\eD$ in \eqref{eq:D-map} defines in turn a trilinear product
\begin{equation}
  \label{eq:3-product}
  \begin{aligned}[m]
    V \times V^* \times V &\to V\\
    (v,\alpha,w) &\mapsto \eD(v,\alpha) \cdot w~.
  \end{aligned}
\end{equation}
Let $\Omega \in (V\otimes V^*)^{\otimes 2}$ be the tensor defined by
\begin{equation}
  \label{eq:Omega}
  \Omega(v,\alpha,w,\beta) := \left<\eD(v,\alpha)\cdot w, \beta\right>~.
\end{equation}

\begin{lemma}\label{le:OmegaSym}
  For all $\alpha,\beta\in V^*$ and $v,w\in V$,
  \begin{equation*}
    \Omega(v,\alpha,w,\beta)  = \Omega(w,\beta,v,\alpha)~.
  \end{equation*}
  In other words, $\Omega \in S^2(V\otimes V^*)$.
\end{lemma}

\begin{proof}
  This follows from the observation that
  \begin{equation*}
    \Omega(v,\alpha,w,\beta)  = \left(\eD(v,\alpha),\eD(w,\beta)\right)~,
  \end{equation*}
  which is manifestly symmetric.
\end{proof}

There is a converse to this result.

\begin{proposition}\label{pr:reconstruction}
  Let $V$ be a finite-dimensional vector space with dual space $V^*$ and let $\Phi: V \times V^* \times V \to V$ be a trilinear product, defining a
  bilinear map $\eD: V \times V^* \to \End V$ by $\Phi(v,\alpha,w) = \eD(v,\alpha)\cdot w$.  If
  \begin{equation}\label{eq:[D,D]}
    [\eD(v,\alpha),\eD(w,\beta)] = \eD(\eD(v,\alpha)\cdot w,\beta) + \eD(w, \eD(v,\alpha)\cdot \beta)~,
  \end{equation}
  where
  \begin{equation}
    \label{eq:dualmodule}
    \left<w, \eD(v,\alpha)\cdot \beta\right> = -\left<\eD(v,\alpha)\cdot w, \beta\right>~,
  \end{equation}
  and
  \begin{equation}\label{eq:Dsymm}
    \left<\eD(v,\alpha)\cdot w,\beta\right> = \left<\eD(w,\beta)\cdot v,\alpha\right>~,
  \end{equation}
  then $\eD$ is obtained from $(\fg,V)$ by the above construction, where $\fg < \End V$ is the image of $\eD$.
\end{proposition}

\begin{proof}
  Equation~\eqref{eq:[D,D]} says that the image $\fg$, say, of $\eD$ in $\End V$ is a Lie subalgebra making $V$ into a $\fg$-module.
  Equation~\eqref{eq:dualmodule} says that $V^*$ is the dual module.  Define a bilinear form on $\fg$ by extending
  \begin{equation*}
    \left(\eD(v,\alpha),\eD(w,\beta)\right) = \left<\eD(v,\alpha)\cdot w,\beta\right>
  \end{equation*}
  bilinearly to all of $\fg$.  Equation~\eqref{eq:Dsymm} says that it is symmetric.  If $X = \sum_i \eD(v_i,\alpha_i)$ is such that
  $\left(X,\eD(w,\beta)\right)=0$ for all $w\in V$ and $\beta\in V^*$, then it follows as above that $X=0$, whence the inner product on $\fg$ is
  nondegenerate.  Finally we must show that it is $\fg$-invariant, namely, for all $v,w,x\in V$ and $\alpha,\beta,\gamma\in V^*$,
  \begin{equation}
    \label{eq:invariance}
    \left([\eD(v,\alpha),\eD(w,\beta)],\eD(x,\gamma)\right) = \left(\eD(v,\alpha),[\eD(w,\beta),\eD(x,\gamma)]\right)~.
  \end{equation}
  Using equation~\eqref{eq:[D,D]} on the right-hand side, we find
  \begin{align*}
    \left(\eD(v,\alpha),[\eD(w,\beta),\eD(x,\gamma)]\right) &= \left(\eD(v,\alpha),\eD(\eD(w,\beta)\cdot x,\gamma) + \eD(x,\eD(w,\beta)\cdot
      \gamma)\right)\\
    &= \left<\eD(v,\alpha) \cdot \eD(w,\beta)\cdot x,\gamma\right> + \left<\eD(v,\alpha)\cdot x, \eD(w,\beta)\cdot \gamma\right>\\           
    &= \left<\eD(v,\alpha) \cdot \eD(w,\beta)\cdot x,\gamma\right> - \left<\eD(w,\beta)\cdot \eD(v,\alpha)\cdot x, \gamma\right>\\
    &= \left<[\eD(v,\alpha), \eD(w,\beta)] \cdot x,\gamma\right>\\
    &= \left([\eD(v,\alpha), \eD(w,\beta)], \eD(x,\gamma)\right)~,
  \end{align*}
  where in the third equality we have used equation~\eqref{eq:dualmodule}.
\end{proof}

In \cite{FaulknerIdeals} it is shown that many 3-algebras and triple systems can be obtained via this construction, among them (metric) Lie and
Jordan triple systems.  The results in this paper indicate that we can add the hermitian 3-algebras of \cite{BL4} and their generalisations to
them.

\subsection{Embedding Lie (super)algebras}
\label{sec:embedding-general}

It is a natural question to ask, given Remark~\ref{re:KSA}, under which conditions we have a Lie (super)algebra structure on $\fg \oplus V \oplus V^*$
defined by the natural actions of $\fg$ on $V$ and $V^*$ and by the map $\eD: V \times V^* \to \fg$, understood as a Lie bracket.  In this section we
will begin to answer this question and set the stage for future applications of this idea below.

Let us define a $\ZZ_2$-graded vector space $\fk = \fk_0 \oplus \fk_1$ with $\fk_0 = \fg$ and $\fk_1 = V \oplus V^*$ and let us extend the Lie bracket
on $\fk_0$ to a graded bracket
\begin{equation}\label{eq:protoLie}
  [X,v] = X \cdot v \qquad [X,\alpha] = X \cdot \alpha \qquad\text{and}\qquad [v,\alpha] = \eD(v,\alpha)~,
\end{equation}
for all $X\in\fg$, $v\in V$ and $\alpha \in V^*$.  The bracket $[\alpha,v]$ is related to $[v,\alpha]$ by a sign which depends on whether we are trying
to obtain a Lie superalgebra or a $\ZZ_2$-graded Lie algebra.

\begin{remark}\label{re:3grading}
  In fact, the brackets \eqref{eq:protoLie} respect the 3-gradation $V^* \oplus \fg \oplus V$, which refines the 2-gradation and can be useful to check
  the Jacobi identities.
\end{remark}

We would like to understand the conditions guaranteeing that the brackets \eqref{eq:protoLie} define a Lie (super)algebra on $\fk$.  This amounts to
imposing the relevant Jacobi identities.  Being $\ZZ_2$-graded, there are four components to the Jacobi identity.  The $000$-Jacobi is satisfied by
virtue of $\fg$ being a Lie algebra, whereas the $001$-Jacobi follows from the fact that $V$ and $V^*$ are $\fg$-modules.  The $011$-Jacobi is
tantamount to the invariance of $\eD$ under $\fg$, which is precisely what equation~\eqref{eq:ideal} says.  We are left, as usual in these situations,
with checking the $111$-Jacobi.  There are two such Jacobi identities, each with a choice of sign depending on whether we are after a Lie superalgebra
or a $\ZZ_2$-graded Lie algebra:
\begin{equation}
  \label{eq:111Jacobi}
  \begin{aligned}[m]
    \eD(v,\alpha) \cdot  w &= \pm \eD(w,\alpha) \cdot v\\
    \eD(v,\alpha) \cdot \beta &= \pm \eD(v,\beta) \cdot \alpha~,
  \end{aligned}
\end{equation}
for all $v,w\in V$ and $\alpha,\beta\in V^*$, where the upper sign is for the case of a $\ZZ_2$-graded Lie algebra and bottom sign for the case of a Lie
superalgebra.  Finally, it is not hard to see that both identities are equivalent.  Indeed the first equation in \eqref{eq:111Jacobi} says that
$\Omega(v,\alpha,w,\beta) := \left<\eD(v,\alpha)\cdot w, \beta\right>$ defined in \eqref{eq:Omega} is (skew)symmetric in $(u,v)$, whereas the second
equation says that it is (skew)symmetric in $(\alpha,\beta)$.  However Lemma \ref{le:OmegaSym} says that these two conditions are equivalent.

\subsection{Unitary representations}
\label{sec:unitary-modules}

We will now briefly consider a special case of the above construction, in which $V$ possesses an inner product invariant under $\fg$ and in particular
under the $\eD(v,\alpha)$.  We will consider only those kinds of inner products which have a notion of signature, since in order to build lagrangians for
manifestly unitary theories, we will want to consider the positive-definite ones.  There are three such inner products: real symmetric, complex
hermitian and quaternionic hermitian.  In all cases, the inner product allows us to identify $V$ and the dual $V^*$; although in the complex hermitian
case this identification is complex antilinear and hence $V$ and $V^*$ are inequivalent as representations of $\fg$.  This means that
equation~\eqref{eq:dualmodule}, once $V$ and $V^*$ have been identified, becomes one more condition on $\eD(v,\alpha)$.

\subsubsection{Real orthogonal representations}
\label{sec:real-metric}

Here we will take $V$ to be a faithful real orthogonal representation of $\fg$.  The inner product on $V$ defines in particular musical isomorphisms
$\flat: V \to V^*$ and $\sharp: V^* \to V$, which are inverses of each other.  The bilinear map $\eD: V \times V^* \to \fg$ of equation \eqref{eq:D-map}
gives rise to a bilinear map $D: V \times V \to \fg$, defined by $(u,v)\mapsto D(u,v) = \eD(u,v^\flat)$ and viceversa with $\eD(v,\alpha) =
D(v,\alpha^\sharp)$.  Under this dictionary, the construction in Section \ref{sec:Faulkner} coincides with the construction in Section
\ref{sec:reconstruction-CS} and hence the resulting 3-algebras are the generalised metric Lie 3-algebras of Definition~\ref{def:CS}; although they are
contained already in Example~II of \cite{FaulknerIdeals}.

\subsubsection{Complex unitary representations}
\label{sec:hermitian-C}

Here we take $(V,h)$ to be a faithful complex unitary representation of $\fg$.  The musical maps
\begin{equation*}
  \begin{aligned}[m]
    V &\to V^*\\
    v &\mapsto \hfl{v}
  \end{aligned}
  \qquad\text{and its inverse}\qquad
  \begin{aligned}[m]
    V^* &\to V\\
    \alpha &\mapsto \hsh{\alpha}
  \end{aligned}
\end{equation*}
where $\hfl{v}(w) = h(w,v)$ are now complex antilinear, hence they do not provide an isomorphism of the representations $V$ and $V^*$ but rather of
$V^*$ with the complex-conjugate representation $\Vbar$.  Furthermore, since $V$ is a complex vector space, the bilinear map $\eD$ of equation
\eqref{eq:D-map} now maps to the complexification $\fg_\CC = \CC \otimes_\RR \fg$ of $\fg$.  Using the inner product, this map $\eD : V \times V^* \to
\fg_\CC$ gives rise to a sesquilinear map $D: V \times V \to \fg_\CC$, defined by $(u,v) \mapsto D(u,v) = \eD(u,\hfl{v})$ and viceversa with
$\eD(v,\alpha) = D(v,\hsh{\alpha})$.  Under this dictionary, the construction in Section \ref{sec:Faulkner} coincides with the construction in Section
\ref{sec:reconstruction-BL4} and hence the resulting 3-algebras are the hermitian 3-algebras obtained in that section.

\subsubsection{Quaternionic unitary representations}
\label{sec:hermitian-H}

Finally we consider the case where $V$ is a quaternionic unitary representation of $\fg$, by which we will mean a complex unitary representation with a
compatible $\fg$-invariant quaternionic structure.  More precisely, let $(V,h)$ be an (even-dimensional) complex hermitian vector space and let $J: V
\to V$ be a complex antilinear map satisfying $J^2 = - \id$ and
\begin{equation}
  \label{eq:H-struct}
  h(Jx, Jy) = h(y,x)~.
\end{equation}
Moreover both $h$ and $J$ are invariant under $\fg$.  Using $h$ and $J$ we can construct a $\fg$-invariant complex symplectic structure
\begin{equation}
  \label{eq:C-symplectic}
  \omega(x,y) := h(x,Jy)~.
\end{equation}
It is an easy exercise to show that $\omega$ is indeed complex bilinear and that $\omega(x,y) = -\omega(y,x)$.  Since $\omega$ is clearly
$\fg$-invariant, it provides a $\fg$-equivariant symplectic musical isomorphism $\flat: V \to V^*$ with inverse $\sharp: V^* \to V$.  In particular, as
representations of $\fg$, $V$ and $V^*$ are equivalent.

The bilinear map $\eD$ of equation \eqref{eq:D-map} again maps to the complexification $\fg_\CC = \CC \otimes_\RR \fg$ of $\fg$.  We will make $\fg_\CC$
into a complex metric Lie algebra by extending the bracket and inner product complex bilinearly.  As before we will extend also the action of $\fg$ on
$V$ to an action of $\fg_\CC$.  Since $\omega$ is complex bilinear, it remains invariant under $\fg_\CC$.  Using $\omega$, the map $\eD : V \times V^*
\to \fg_\CC$ gives rise to a complex bilinear map $D_\omega: V \times V \to \fg_\CC$, defined by $(u,v) \mapsto D_\omega(u,v) = \eD(u,v^\flat)$ and
viceversa with $\eD(v,\alpha) = D_\omega(v,\alpha^\sharp)$.  Because $(V,h)$ is a complex hermitian vector space on which $\fg$ acts unitarily, the
discussion in Section~\ref{sec:hermitian-C} applies.  In particular, the hermitian structure $h$ defines a sesquilinear map, which we now denote $D_h: V
\times V \to \fg_\CC$ in order to distinguish it from $D_\omega$, by $D_h(u,v) = \eD(u,\hfl{v})$.  The two maps $D_\omega$ and $D_h$ are related as
follows.

\begin{lemma}\label{le:C-H-dict}
  For all $u,v\in V$, $D_\omega(u,v) = D_h(u,Jv)$.
\end{lemma}

\begin{proof}
  Given $u,v\in V$ and for all $\XX \in \fg_\CC$ we have
  \begin{equation}
    \label{eq:D-H}
    \left(D_\omega(u,v), \XX\right) = \omega(\XX \cdot u, v)~.
  \end{equation}
  Using equation \eqref{eq:C-symplectic}, we can rewrite $\omega(\XX \cdot u, v)  = h(\XX \cdot u, Jv)$, whence using equation \eqref{eq:D-C} for $D_h$,
  we obtain
  \begin{equation*}
    \left(D_\omega(u,v), \XX\right)  = h(\XX \cdot u, Jv) = \left(D_h(u,Jv), \XX\right)~.
  \end{equation*}
  Since this is true for all $\XX$, the lemma follows.
\end{proof}

The first thing we notice is that in contrast with the complex hermitian case, $D_\omega(u,v)$ is now symmetric in its arguments.  Indeed,
\begin{align*}
  \left(D_\omega(u,v), \XX \right) &= \omega (\XX \cdot u, v) &&\text{by \eqref{eq:D-H}}\\
  &= -\omega(u, \XX \cdot v) &&\text{since $\fg_\CC$ preserves $\omega$}\\
  &= \omega(\XX\cdot v, u) && \text{since $\omega$ is skewsymmetric}\\
  &= \left(D_\omega(v,u),\XX\right) &&\text{again by \eqref{eq:D-H}.}
\end{align*}
In other words, $D_\omega: S^2V \to \fg_\CC$ and hence defines a complex linear 3-bracket $S^2 V \otimes V \to V$ by $(x,y,z) \mapsto [x,y,z]:=
D_\omega(x,y)\cdot z$, which is symmetric in the first two arguments
\begin{equation}
  \label{eq:symm-H}
  [x,y,z] = [y,x,z]\qquad\text{for all $x,y,z\in V$}.
\end{equation}
We may decompose $S^2V \otimes V$ as
\begin{equation*}
  S^2 V \otimes V = S^3 V \oplus V^{\yng(2,1)}~,
\end{equation*}
where $V^{\yng(2,1)}$ contains the symplectic trace and is therefore not irreducible.  Analogous to the case of generalised metric Lie 3-algebras, there
are two extreme cases of these algebras: namely those for which the 3-bracket is totally symmetric and those for which it obeys the condition
\begin{equation}
  \label{eq:aLTSc}
  [x,y,z] + [y,z,x] + [z,x,y] = 0~.
\end{equation}
These latter triple systems are called \textbf{anti-Lie triple systems} and are discussed in \cite[§5]{FaulknerFerrarAJP}. Just like a Lie triple system
can be embedded in a Lie algebra, an anti-Lie triple system can be embedded in a Lie superalgebra.  Indeed on the $\ZZ_2$-graded vector space $\fg_\CC
\oplus V$, where $\fg_\CC$ has even parity and $V$ odd parity, we can define the following brackets extending those of $\fg_\CC$:
\begin{equation*}
  [D_\omega(x,y), z] = [x,y,z]\qquad\text{and}\qquad [x,y] = D_\omega(x,y)~.
\end{equation*}
Since $D_\omega(x,y) = D_\omega(y,x)$ we see that the bracket on $V$ is indeed symmetric.  As usual there are four components to the Jacobi identity, of
which only the component in $S^3 V \to V$ needs to be checked.  This component of the Jacobi identity requires $[x,[x,x]]=0$ for all $x\in V$, which is
equivalent to $[x,x,x]=0$, which in turn is equivalent (using the symmetry of $D_\omega$) to the defining condition \eqref{eq:aLTS}.

Furthermore the complex Lie superalgebra $\fg_\CC \oplus V$ is metric.  We simply use the inner product on $\fg_\CC$ and the $\fg_\CC$-invariant
symplectic form $\omega$ on $V$, while declaring $\fg_\CC \perp V$ as befits an even inner product.  Both are $\fg_\CC$-invariant, hence the only
compatibility condition is precisely equation \eqref{eq:D-H}.  Notice however that $V$ is not a real representation of $\fg$, whence this complex Lie
superalgebra is not the complexification of a real Lie superalgebra.

Three remarks are in order.  The first remark is that the emergence of a Lie superalgebra here is a different phenomenon than that of a Lie superalgebra
in the hermitian 3-algebras of \cite{BL4} discussed in Section~\ref{sec:recognition-BL4}.  First of all, the underlying space of the Lie superalgebra is
different, but moreover the condition for the existence of an embedding superalgebra in that case was equation \eqref{eq:skew-BL4}, which in the
notation of this section becomes $D_h(x,y)\cdot z = - D_h(z,y)\cdot x$.  In terms of $D_\omega$, this condition says that the 3-bracket $[x,y,z] =
D_\omega(x,y)\cdot z$ obeys
\begin{equation}
  \label{eq:skew-BL4-H}
  [x,y,z] = - [x,z,y]~.
\end{equation}
However this is inconsistent with the symmetry condition $[x,y,z] = [y,x,z]$ unless the 3-bracket and hence $D$ vanishes identically; indeed,
\begin{equation*}
  [x,y,z] \stackrel{\eqref{eq:symm-H}}{=} [y,x,z] \stackrel{\eqref{eq:skew-BL4-H}}{=} - [y,z,x] \stackrel{\eqref{eq:symm-H}}{=} - [z,y,x]
  \stackrel{\eqref{eq:skew-BL4-H}}{=} [z,x,y] \stackrel{\eqref{eq:symm-H}}{=} [x,z,y] \stackrel{\eqref{eq:skew-BL4-H}}{=} - [x,y,z]~.
\end{equation*}

The second remark is that the link between Lie superalgebras and triple systems is itself not surprising.  Any 2-graded algebra (e.g., a Lie
superalgebra) defines a triple system on the elements of odd parity, simply by nesting the product on the algebra: $(x,y,z) \mapsto [[xy]z]$, for all
$x,y,z$ odd elements and where $[xy]$ denotes the multiplication in the 2-graded algebra.  Which type of triple system depends on which type of 2-graded
algebra we employ.  For example, we have seen that we can obtain Lie triple systems from 2-graded Lie algebras and hermitian 3-algebras and anti-Lie
triple systems starting from certain kinds of Lie superalgebras.  This is done explicitly for exceptional Lie superalgebras by Okubo \cite{Okubo:2003uh}
in terms of so-called \emph{$(-1,-1)$-balanced Freudenthal--Kantor} triple system.

The third and last remark is that, of course, not every anti-Lie triple system is of this form, since remember that $V$ possesses a $\fg$-invariant
quaternionic structure $J$, which we will now exploit more fully.

The existence of the quaternionic structure is equivalent to the fact that this 3-algebra is a special case of the hermitian 3-algebras of
Section~\ref{sec:hermitian-C}.  This allows us to deduce some new identities by relating the properties of $D_\omega$ and of $D_h$ via the dictionary in
Lemma~\ref{le:C-H-dict}.  In particular, since $V$ is a faithful representation, $D_\omega$ is surjective onto $\fg_\CC$ since so is $D_h$.

Since $J$ is $\fg$-invariant but complex antilinear, it follows that
\begin{equation}
  \label{eq:D-J-H}
  D_\omega(x,y) \circ J = J \circ \overline{D_\omega(x,y)}~.
\end{equation}
Similarly, the identity \eqref{eq:Dbar} satisfied by $D_h$ becomes
\begin{equation}
  \label{eq:Dbar-H}
  D_\omega(x,Jy) = - \overline{D_\omega(y,Jx)}~,
\end{equation}
which may be rewritten as
\begin{equation}
  \label{eq:bracket-J}
  J[x,y,z] = [Jx, Jy, Jz]
\end{equation}
with the help of equation \eqref{eq:D-J-H}.

Using the two conditions \eqref{eq:D-J-H} and \eqref{eq:Dbar-H}, it is not hard to show that the identities \eqref{eq:unitarity-C} and
\eqref{eq:symmetry-C}, when written in terms of $D_\omega$, are automatically satisfied by virtue of the symmetry properties satisfied by the
fourth-rank tensor
\begin{equation*}
  \Theta(x,y,z,w) :=  \omega([x,y,z],w)~.
\end{equation*}
These properties, which follow at once from \eqref{eq:symm-H} and from the fact that $\Theta(x,y,z,w) = \left(D_\omega(x,y), D_\omega(z,w)\right)$, are
seen to be following:
\begin{equation}
  \label{eq:symmetry-H}
  \Theta(x,y,z,w) = \Theta(y,x,z,w) = \Theta(z,w,x,y) = \Theta(x,y,w,z)~,
\end{equation}
whence $\Theta \in S^2(S^2 V^*)$.

Finally, we come to the fundamental identity.  This can be done in any number of ways.  One way is to use Lemma \ref{le:C-H-dict} to relate the
3-brackets to those of the underlying hermitian 3-algebra as follows:
\begin{equation*}
  [x,y,z] = D_\omega(x,y) \cdot z = D_h(x, J y) \cdot z = [z,x; J y]~,
\end{equation*}
and then rewrite the fundamental identity \eqref{eq:FI-C} of the \emph{general} hermitian 3-algebra in terms of the 3-brackets $[x,y,z]$.
Alternatively, we can simply apply equation \eqref{eq:[D,D]} to $z\in V$, say, and express the result in terms of the 3-brackets $[x,y,z]$.  Either way,
we arrive at the following \emph{fundamental identity}
\begin{equation}
  \label{eq:FI-H}
  [v,x,[w,y,z]] - [w,y,[v,x,z]] - [[v,x,w],y,z] - [w,[v,x,y],z] = 0~.
\end{equation}

In summary, the most general 3-algebra constructed out of a metric Lie algebra $\fg$ and a faithful quaternionic representation $(V,h,J)$ is defined by
a complex trilinear 3-bracket $V\times V \times V \to V$, satisfying the following minimal set of axioms: \eqref{eq:symm-H}, \eqref{eq:bracket-J},
\eqref{eq:FI-H} and
\begin{equation}
  \label{eq:symmetry-H-2}
  \omega([x,y,z],w) = \omega([z,w,x],y)~,
\end{equation}
which is one of the relations in equation \eqref{eq:symmetry-H} when written explicitly in terms of the 3-bracket.

The simple classical Lie superalgebras of types $B(m,n)$ and $D(m,n)$ can be seen to arise in this way, as the following example shows.  This can be
thought of as the quaternionic analogue of Example \ref{eg:BL4}.

\begin{example}\label{eq:aLTS}
  Let $\fg = \fso(m) \oplus \fusp(2n)$.  We let $U$ be the complexification of the fundamental representation of $\fso(m)$; that is, $U \cong \CC^m$
  with an invariant real structure or, equivalently, an invariant complex symmetric inner product which we will denote $g$.  Let $W$ be the fundamental
  representation of $\fusp(2n)$, so $W \cong \CC^{2n}$ with an invariant quaternionic structure or, equivalently, an invariant complex symplectic
  structure which we will denote $\varepsilon$.  We let $V = U \otimes W$ be the \emph{bifundamental} representation.  Our aim is to define an anti-Lie
  triple system on $V$ embedding into the complex simple Lie superalgebra of B or D types, depending on the parity of $m$: odd or even, respectively.
  Using $\varepsilon$ we can (and will) identify $V$ with the space of complex linear maps $W \to U$: $u \otimes w \in V$ is identified with the linear
  map
  \begin{equation*}
    (u \otimes w) (x) := \varepsilon(w,x) u~.
  \end{equation*}
  The action of $\fg$ on $V$ is the following:
  \begin{equation*}
    (X,Y)\cdot \varphi  = X \circ \varphi  - \varphi  \circ Y~, \qquad\text{for all $(X,Y) \in\fg$ and $\varphi : W \to U$.}
  \end{equation*}
  We extend this action to one of the complexification $\fg_\CC$ of $\fg$.  Given $\varphi : W \to U$ we can define its transpose $\varphi^*$ by
  \begin{equation}
    \label{eq:transpose}
    g(\varphi  w, u) = \varepsilon(w, \varphi^* u)~,
  \end{equation}
  for all $w \in W$ and $u \in U$.  This allows us to define a $\fg_\CC$-invariant complex symplectic structure on $V$ by
  \begin{equation}
    \label{eq:ip}
    \omega(\varphi , \psi) := \Tr_W (\psi^* \circ \varphi) = \Tr_U (\varphi \circ \psi^*)~.
  \end{equation}
  Now given $\varphi ,\psi \in V$ we define $D(\varphi ,\psi ) \in \fg_\CC$ by
  \begin{align*}
    \left( D(\varphi ,\psi ), (\XX,\YY) \right) &= \omega\left((\XX,\YY) \cdot \varphi , \psi \right)\\
    &= \omega\left(\XX \circ \varphi - \varphi \circ \YY, \psi \right) \\
    &= \Tr_U (\XX \circ \varphi \circ \psi^*) - \Tr_W (\psi^* \circ \varphi \circ \YY )~,
  \end{align*}
  where the ad-invariant inner product on $\fg_\CC$ is defined by
  \begin{equation*}
    \left((\XX_1,\YY_1), (\XX_2, \YY_2)\right)= \Tr_U (\XX_1 \circ \XX_2)  - \Tr_W (\YY_1 \circ \YY_2)~.
  \end{equation*}
  Writing $D(\varphi ,\psi ) = (\DD_1, \DD_2)$, with $\DD_1 \in \fso(m)_\CC$ and $\DD_2 \in \fusp(2n)_\CC$, we see that
  \begin{equation*}
    \Tr_U (\DD_1 \circ \XX) - \Tr_W (\DD_2 \circ \YY) = \Tr_U (\XX \circ \varphi \circ \psi^*) - \Tr_W (\psi^* \circ \varphi \circ \YY )~,
  \end{equation*}
  whence $\DD_1 = \pr_1( \varphi \circ \psi^* )$ and $D_2 = \pr_2 (\psi^* \circ \varphi)$, where $\pr_1$ and $\pr_2$ are the orthogonal projections
  \begin{equation*}
    \pr_1 : \fgl(m,\CC) \to \fso(m)_\CC \qquad\text{and}\qquad \pr_2 : \fgl(2n,\CC) \to \fusp(2n)_\CC~.
  \end{equation*}
  A simple calculation using equation \eqref{eq:transpose} shows that the transposes of $\varphi \circ \psi^*$ and $\psi^* \circ \varphi$ relative to
  the inner products on $U$ and $W$, respectively, are given by
  \begin{equation*}
    (\varphi \circ \psi^*)^t = - \psi \circ \varphi^*  \qquad\text{and}\qquad (\psi^* \circ \varphi )^t = - \varphi^* \circ \psi~,
  \end{equation*}
  where the signs are due to the skewsymmetry of $\varepsilon$.  This means that
  \begin{equation*}
    2 \DD_1 = \varphi \circ \psi^* + \psi  \circ \varphi^* \qquad\text{and}\qquad 2 \DD_2 = \psi^* \circ \varphi  + \varphi^* \circ \psi~,
  \end{equation*}
  whence
  \begin{equation*}
    2 D(\varphi ,\psi ) = \left(\varphi  \circ \psi^* + \psi \circ  \varphi^*, \psi^* \circ \varphi  + \varphi^* \circ \psi\right)~.
  \end{equation*}
  We may (and will) get rid of the annoying factor of $2$ by retroactively rescaling the inner product on $\fg_\CC$.  The 3-bracket on $V$ is then given
  by
  \begin{align*}
    [\varphi ,\psi ,\chi] &= D(\varphi ,\psi ) \cdot \chi\\
    &= (\varphi \circ \psi^* + \psi \circ \varphi^*) \circ \chi - \chi \circ (\psi^* \circ \varphi  + \varphi^* \circ \psi )~.
  \end{align*}
  It obeys all the identities coming from the Faulkner construction applied to a quaternionic unitary representation and, in addition,
  \begin{equation*}
    [\varphi ,\varphi ,\varphi ] = 0 \qquad\text{for all $\varphi \in V$,}
  \end{equation*}
  whence it is an anti-Lie triple system.  This means that it embeds in a complex Lie superalgebra on the superspace $\fg_\CC \oplus V$.  There is
  precisely one Lie superalgebra with this data---namely the complex orthosymplectic Lie superalgebra, which is simple and of type $B((m-1)/2,n)$, if
  $m$ is odd, and $D(m/2,n)$ is $m$ is even.
\end{example}

These Lie superalgebras were found in \cite{3Lee} to play a rôle in $N{=}5$ superconformal Chern--Simons theories analogous to the rôle played by the
$A(m,n)$ and $C(n+1)$ superalgebras in $N{=}6$ theories.  In addition, in \cite{BHRSS} further $N{=}5$ theories were constructed by taking a limit of
superconformal gaugings of supergravity theories, associated to the exceptional Lie superalgebras $F(4)$ and $G(3)$, as well as to the family of simple
Lie superalgebras $D(2,1;\alpha)$.  These Lie superalgebras too can be constructed along the lines of the previous example.  We will not give the
details, but point the reader to the paper \cite{MR1961174} by Kamiya and Okubo, where the construction is made manifest.  In that paper it is shown
that these Lie superalgebras arise from anti-Lie triple systems which can be constructed from Faulkner data $(\fg, V)$ with $V$ a quaternionic unitary
representation of the real metric Lie algebra $\fg$.  To be precise, for $F(4)$, $\fg = \fso(7) \oplus \fusp(2)$ and $V = S \otimes W$, with $W$ the
complex two-dimensional representation of $W$ with its invariant quaternionic structure and $S$ the complexification of the 8-dimensional irreducible
spinor representation of $\fso(7)$, and for $G(3)$, $\fg = \fg_2 \oplus \fusp(2)$ and $V = E \otimes W$, with $W$ as before and $E$ the complexification
of the ``fundamental'' seven-dimensional irreducible representation of $\fg_2$.  In both of these cases $\fso(7)$ and $\fg_2$ are compact real forms and
the metric structure is provided by (a multiple of) the Killing form.  Finally, for $D(2,1;\alpha)$, $\fg = \fso(4) \oplus \fusp(2)$, and $V = F \otimes
W$, with $W$ as before and $F$ the complexification of the fundamental representation.  The parameter $\alpha$ is reflected in the choice of inner
product for $\fso(4)$, which not being simple has a two-parameter family of invariant inner products.  Up to an overall rescaling of the inner product
we are left with a pencil of such inner products, parametrised rationally by $\alpha$.

From these examples we conclude that all the embedding superalgebras involved in the new $N=5$ superconformal Chern-Simons-matter theories appearing in
the recent literature \cite{3Lee,BHRSS} can also be realised in the quaternionic hermitian case of the Faulkner construction.

\section{Conclusions and open problems}
\label{sec:conclusions}

In this paper we have hopefully shed some light on the Lie-algebraic origin of the metric 3-algebras which lurk behind three-dimensional superconformal
Chern--Simons theories.  In particular, we have proved that the metric 3-algebras of \cite{CherSaem}, appearing in $N{=}2$ theories, correspond to pairs
$(\fg, V)$ consisting of a metric Lie algebra $\fg$ and a real faithful orthogonal representation $V$.  We have also proved that a class of hermitian
3-algebras generalising those in \cite{BL4} correspond to pairs $(\fg, V)$ where $\fg$ is again a metric Lie algebra and $V$ is a faithful complex unitary
representation.  The class of hermitian 3-algebras in \cite{BL4}, which appear in the $N{=}6$ theories, constitute a subclass which are in one-to-one
correspondence with a class of metric Lie superalgebras into whose complexification they embed.  We have shown that these constructions are special
cases of a general construction of pairs due to Faulkner.  Applying this to the case of pairs $(\fg, V)$ where $V$ is now a quaternionic unitary
representation of the real metric Lie algebra $\fg$, we obtain a number of complex 3-algebras generalising those which arise in $N{=}5$ superconformal
Chern--Simons theories \cite{3Lee,BHRSS}.

These results begin to explain, in simple algebraic terms, the fact that superconformal Chern--Simons theories which were originally formulated in terms of 3-algebras, can be rewritten using only Lie algebraic data, as in standard gauge theories (albeit with a twist).  They similarly
explain the relation between Lie superalgebras and $N\geq 4$ superconformal Chern--Simons theories.  The repackaging of 3-algebraic data in Lie
algebraic terms is very compact and has the advantage of being able to construct examples very easily.  On the other hand, the very efficiency of this
repackaging makes it hard to rephrase properties of the 3-algebra in terms of Lie algebras.  Along these lines, there a number of open problems and
directions for future research, some of which we are busy exploring and will be reported on elsewhere, which are suggested by the results presented
here.

The dictionary between metric 3-algebras and Lie-algebraic data presented here points at the possibility of rephrasing properties of the 3-algebra (and
of the associated superconformal Chern--Simons theory) directly in terms of the pairs $(\fg, V)$.  The structure theory of metric 3-algebras, which for
the case of metric Lie 3-algebras was exploited in \cite{Lor3Lie,2p3Lie} in order to classify those algebras of index $<3$ and rephrase properties of
the Bagger--Lambert model in 3-algebraic terms, ought to be encoded in Lie-algebraic terms.  An important open problem in this regard is how to
recognise the type of metric 3-algebra from the Lie-algebraic data.  In particular, which $(\fg,V)$ give rise to metric Lie 3-algebras?

The construction of superconformal Chern--Simons theories directly from $(\fg,V)$ is an interesting problem on which we will report in a forthcoming
paper.  A better understanding of the dictionary between 3-algebras underlying the $N{=}3$ theories dual to 3-Sasaki 7-manifolds (and the $N{=}1$
theories associated to their squashings) and the geometry of these manifolds would also be desirable.  In particular, how does the geometry of the
3-Sasaki manifold (or its squashing) manifest itself in its 3-algebra or in its Lie-algebraic data?

\section*{Acknowledgments}

JMF would like to extend his gratitude to José de Azcárraga and to the Departament de Física Teòrica of the Universitat de València for hospitality and
support under the research grant FIS2008-01980.  PdM is supported by a Seggie-Brown Postdoctoral Fellowship of the School of Mathematics of the
University of Edinburgh.

\bibliographystyle{utphys}
\bibliography{AdS,AdS3,ESYM,Sugra,Geometry,Algebra}

\end{document}